\documentclass[journal ]{new-aiaa}
\usepackage[utf8]{inputenc}
\usepackage{textcomp}

\usepackage{bm}
\usepackage{array}
\usepackage{graphicx}
\usepackage{amsmath}
\usepackage{float}
\usepackage{amssymb}
\usepackage{subfigure}
\usepackage{xcolor,color,colortbl}
\usepackage[version=4]{mhchem}
\usepackage{siunitx}
\usepackage{longtable,tabularx}
\usepackage{comment}
\title{Reinforcement Meta-Learning for Interception of Maneuvering Exoatmospheric Targets with Parasitic Attitude Loop}

\author{Brian Gaudet\footnote{Engineer, E-mail:briangaudet@mac.com}}
\affil{DeepAnalytX, LLC, 1130 Swall Meadows Rd,  Bishop CA 93514}
\author{Roberto Furfaro\footnote{Professor, Department of Systems and Industrial Engineering, Department of Aerospace and Mechanical Engineering}}
\affil{University of Arizona, 1127 E. Roger Way, Tucson Arizona, 85721}
\author{Richard Linares\footnote{Charles Stark Draper Assistant Professor, Department of Aeronautics and Astronautics, Senior Member AIAA, E-mail:linaresr@mit.edu}}
\affil{Massachusetts Institute of Technology, Cambridge, MA 02139}
\author{Andrea Scorsoglio\footnote{Phd Candidate}}
\affil{University of Arizona, 1127 E. Roger Way, Tucson Arizona, 85721}

\begin{document}

\maketitle

\begin{abstract}
We use Reinforcement Meta-Learning to optimize an adaptive integrated guidance, navigation, and control system suitable for exoatmospheric interception of a maneuvering target.  The system maps observations consisting of  strapdown seeker angles and rate gyro measurements directly to thruster on-off commands. Using a high fidelity six degree-of-freedom simulator, we demonstrate that the optimized policy can adapt to parasitic effects including seeker angle measurement lag, thruster control lag, the parasitic attitude loop resulting from scale factor errors and Gaussian noise on angle and rotational velocity measurements, and a time varying center of mass caused by fuel consumption and slosh.  Importantly, the optimized policy gives good performance over a wide range of challenging target maneuvers. Unlike previous work that enhances range observability by inducing line of sight oscillations, our system is optimized to use only measurements available from the seeker and rate gyros. Through extensive Monte Carlo simulation of randomized exo-atmospheric interception scenarios, we demonstrate that the optimized policy gives performance close to that of augmented proportional navigation with perfect knowledge of the full engagement state. The optimized system is computationally efficient and requires minimal memory, and should be compatible with today's flight processors.  
\end{abstract}

\section{Introduction}
\lettrine{D}{ue} to the hit to kill requirement and  small size of a ballistic reentry vehicle (BRV), parasitic effects in both the interceptor's internal dynamics and guidance, navigation, and control (GN\&C) system can have a large impact on the probability of a successful intercept. The interception problem is further complicated by warheads with limited maneuvering capability. Both spiral and bang-bang maneuvers could potentially be executed by a BRV without compromising the BRV's accuracy. These maneuvers could be executed either in response to the BRV's sensor input (if so equipped) or periodically executed during the portion of the trajectory where interception is likely. Exoatmospheric interceptors typically use a strapdown infrared (IR) seeker, where the seeker is fixed in the missile body frame with the seeker boresight axis aligned with the missile centerline. This approach has the advantage of simplified design and fewer moving parts as compared to a gimbaled seeker, but does introduce parasitic effects.  

To avoid confusing missile rotations with target maneuvers, a seeker must be stabilized so that  measurements are taken from an inertial reference frame. In a gimbaled seeker arrangement, the seeker platform is adjusted mechanically during the engagement to compensate for missile rotation. However, since a strapdown seeker is fixed in the missile body frame, the seeker angle  measurements are taken in a reference frame that in general is rotating. Consequently, stabilization must be accomplished computationally. The approach most often described in the literature is to subtract rate gyro yaw and pitch rates from the azimuth and elevation seeker angle rates of change \cite{hong2019study,hong2014compensation, willman1988effects, kim2007stability}, although this requires roll stabilization to decouple the yaw and pitch channels.  A parasitic feedback loop that occurs with this approach is due to a delay mismatch between the yaw / pitch rates and the rate of change of the seeker angles \cite{hong2019study,hong2014compensation}.  This can be designed around by adding a delay to align the two signals . Although the delays will in general vary with semiconductor process variations, temperature, and source voltage variations, if the delays are bounded they can be precisely matched  using a delay locked loop \cite{gaudet2000dll,gaudet2003method}.  Another parasitic parasitic feedback loop is described in \cite{willman1988effects,kim2007stability}, and is due to scale factor errors in the rate gyro and seeker measurements. These scale factor errors can cause a false indication of change in the stabilized seeker angles that the guidance system will attempt to nullify; this is described in more detail in Section \ref{PAL}. Borrowing the terminology from \cite{siouris2004missile:1}, we will refer to these parasitic feedback loops as the "parasitic attitude loop", because they are only activated when the missile rotates. The parasitic attitude loop also occurs in endoatmospheric missiles equipped with a gimbaled seeker and radome, with the latter refracting the incoming signal at an angle proportional to the signal's angle with the body-frame axis \cite{zarchan2012tactical:5, siouris2004missile:1}.

A parasitic effect that to our knowledge has been neglected in the literature (although it has been addressed in the context of orbital refueling \cite{guang2018attitude}) is caused by changes in the missile's center of mass during the homing phase. To the extent that the center of mass dynamics results in unwanted missile rotation, this parasitic effect is an integral part of the aforementioned parasitic attitude loop. As the missile fires its divert thrusters to minimize the line of sight (LOS) rate, fuel is consumed, which shifts the missile's center of mass.  Consequently, divert thrusts will in general create  both translational forces and body frame torques, resulting in undesired rotational motion, with the magnitude of the coupling between translational and rotational motion increasing as fuel is consumed. Careful engineering of the fuel system can reduce, but not completely eliminate this effect. The consumption of fuel also reduces the missile's mass and modifies the missile's inertia tensor, with the changing inertia tensor inducing an additional rotational force (See Section \ref{EQOM}).  A related problem is fuel slosh, where the forces created by divert thrusts create oscillations in the missile's center of mass.  Again, the problem can be mitigated by the use of baffles in the fuel storage tanks, but not completely eliminated. Both the unwanted torque induced by divert thrusts and the associated  corrections by the attitude control thrusters induce changes in the missile's rotational velocity and associated changes in the unstabilized body-frame seeker angles, which activates the parasitic attitude loop. We discuss this in more detail in Section \ref{PAL}, where we can present the argument in the context of our seeker computational stabilization method.


Another complication is that since exoatmospheric interceptors use infrared (IR) seekers, range and range rate to target are not available for use in a guidance law. Proportional navigation (PN) requires closing velocity, and augmented proportional navigation (APN) requires an estimate of target acceleration. Optimal control approaches typically require the full engagement state (relative range and velocity, and target acceleration). The work relating to a missile guidance law using angle-only measurements is relatively scarce \cite{taur1999passive, song1996practical, reisner2013optimal, gaudet2020reinforcement}, and most use an approach where the guidance law enhances range observability by deliberately inducing rotations in the LOS angles during the engagement, allowing the range to be inferred using a recursive Bayesian filter, typically a modified Kalman filter. Although in the above cited works the induced rotations did not impact performance, it is not clear that this would be the case in general. To our knowledge, aside from \cite{gaudet2020reinforcement}, there is no published work describing a guidance law that actually maps seeker angles directly to actuator commands without first attempting to infer range to target.

Our goal in this work is to design and optimize a GN\&C system for the terminal phase of an exoatmospheric intercept, with the system allowing the missile to reliably intercept a highly maneuvering target with a miss distance less than 50cm.  The GN\&C system will adapt to both the target maneuver and the potentially time-varying parasitic effects described above. To enhance performance, the system will be fully integrated, mapping sensor outputs (with a minimal amount of processing for computational stabilization) directly to divert and attitude thrust commands, exploiting synergies between the three separate systems \cite{palumbo2004integrated}. Although integration of guidance and control has been extensively covered in the literature \cite{palumbo2004integrated, padhi2014partial,  koren2008integrated, panchal2017continuous}; we could not find any work addressing integrated GN\&C applied to missile guidance. Moreover, these works assumed perfect knowledge of the engagement state, and with the exception of \cite{palumbo2004integrated, padhi2014partial}, these works were not verified in a full 6-DOF simulation. It is worth noting that recent work has explored integrated GN\&C for spaceflight applications \cite{gaudet2019seeker, gaudet2020six}.

We optimize the integrated GN\&C system using  reinforcement meta-learning (Meta-RL)\footnote{See Section \ref{RL} for an overview of  Meta-RL}. The system maps seeker angles and missile rotational velocity directly to divert and attitude thrust commands, and respects seeker field of view and maximum thrust constraints. Learning involves having an agent instantiating the system interact episodically with an environment. Each episode consists of an engagement scenario with randomized parameters.  Here we consider a maneuverable BRV high-altitude terminal phase interception scenario, where the intercepting missile must destroy the target kinetically via a direct hit (miss less than 50 cm). The details of this engagement scenario are provided in Section \ref{engagement}, with the environmental dynamics given in Section \ref{EQOM}. Importantly, the Meta-RL policy instantiated in the system is adaptive, in that after a few steps of interaction with the environment, the policy will be able to adapt to parasitic effects that in general may vary between episodes. This work differs from previous published work as described below:

\begin{enumerate}
\item \underline{Fully Integrated GN\&C:} The system maps seeker angles and missile rotational velocity directly to divert and attitude thrust commands. Although one might argue that this system is not fully integrated in that we computationally stabilize the body frame seeker angles and pass them through a low pass filter for smoothing, our approach does not require estimation of the full engagement state, as is the case with optimal control and sliding mode based systems, or even range estimation.
\item \underline{High Fidelity Simulator:} Full 6-DOF simulation including modeling of parasitic effects including the parasitic attitude loop, changes in inertia tensor and center of mass due to fuel consumption, actuator lag, and seeker angle filtering lag. While this still leaves out other significant parasitic effects and noise components, we believe we capture the essence of the problem, and could replicate these results with a higher fidelity environment.
\item \underline{Highly Maneuvering Target}: The target has acceleration capability equal to 1/2 that of the missile, and executes maneuvers including bang-bang, vertical-S and barrel roll.  Since the BRV must carry a nuclear device, we would expect the interceptor to have the advantage in maneuverability. 
\item \underline{Target Tracking Approach}: Rather than assume that the focal plane array has a narrow field of view (FOV), which requires keeping the seeker pointed at the target using attitude control thrusters, we take advantage of the next generation of focal plane arrays, which will allow 90 degree or higher FOV, allowing the missile to remain pointed at where the target is likely to be at the time of intercept. This gives many performance advantages, as  we describe in Section \ref{seeker}.
\item \underline{Novel Attitude Stabilization Scheme }:  We present an obvious, but to our knowledge unpublished, attitude stabilization scheme. Rather than assume decoupled yaw and pitch channels to subtract rotational velocity directly from seeker angle rate of change, we rotate the seeker angles back to the inertial frame associated with the missile attitude at the start of the maneuver, integrating the rotational velocity to provide an estimate of the change in attitude. Being fully general, this works with coupled yaw and pitch channels, and we demonstrate that it is reasonably robust to scale factor errors and Gaussian noise.
\item \underline{Guidance Frequency:}  We use a guidance frequency of 25 Hz, whereas most published 6-DOF work for exoatmospheric intercept uses a 100 Hz frequency. Assuming that the rate gyros and focal plane array can provide measurements at 100 Mhz, this gives four cycles that could productively be used to filter noise.
\end{enumerate}

A system diagram illustrating the interface between the GN\&C system modeled in this work and peripheral system components is shown below in Figure \ref{fig:System}. In addition, the figure illustrates the internal interface between the modeled GN\&C subsystems and the Meta-RL policy. Target discrimination is not implemented in this work, and we assume that the target discrimination block can return the (noisy) body-frame seeker angles for the actual target. The details of the integrator and computational stabilization scheme are given in Section \ref{seeker}.  The system optimization and testing environment includes the full GN\&C system, noise models, and a thruster model. 

\begin{figure}[h]
\begin{center}
\includegraphics[width=1.0\linewidth]{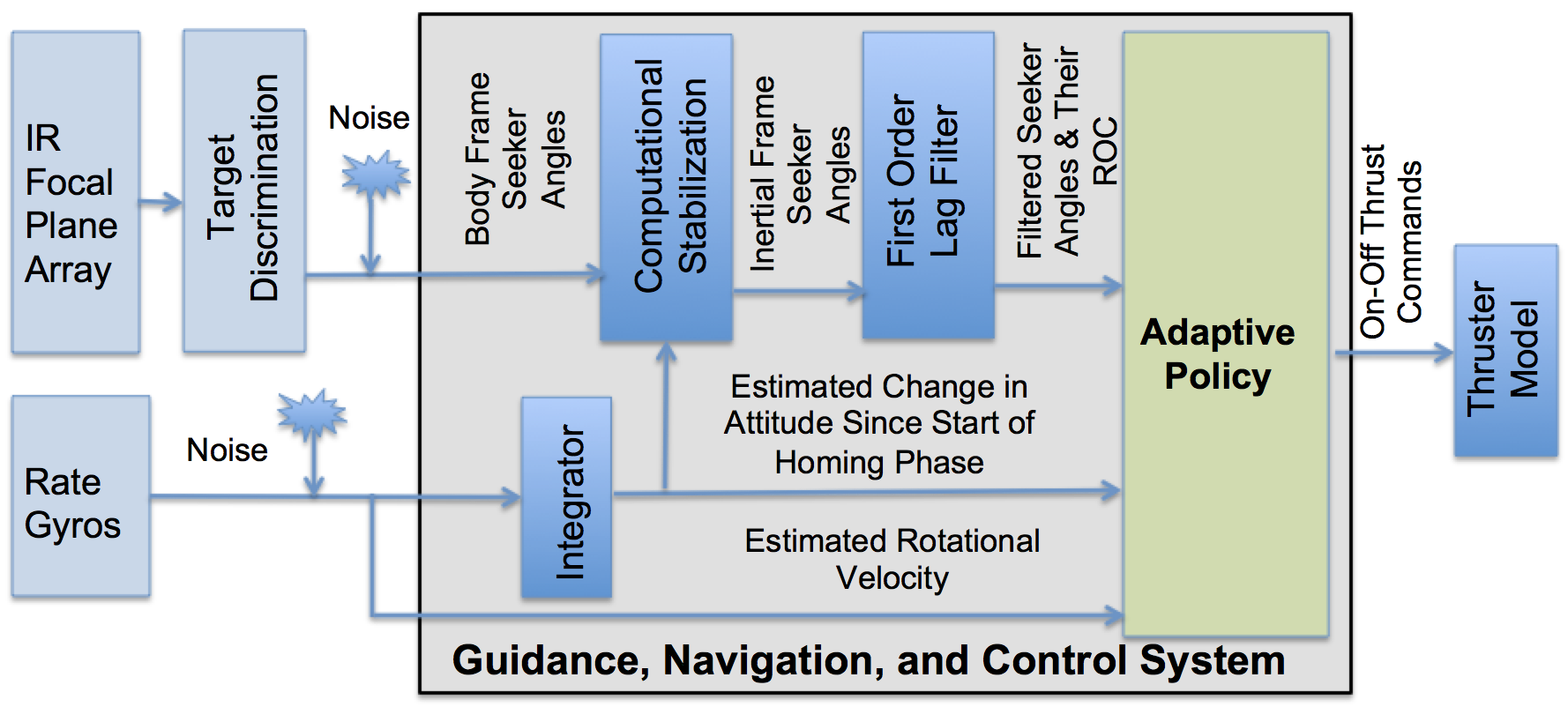}
\caption{Engagement}
\label{fig:System}
\end{center}
\end{figure}

As a benchmark, we modify the augmented proportional navigation (APN) guidance law \cite{zarchan2012tactical:5} to work with on/off thrusters. The APN guidance law is tested against the same engagement scenarios as the Meta-RL policy, but in a 3-DOF simulator, where the guidance law is not confronted with the parasitic effects described earlier. In Section \ref{Experiments}, we test our GN\&C system in a 6-DOF simulator that models parasitic effects, and demonstrate that the system's performance is close to that of the APN guidance law.  Since coupling the APN guidance law with a navigation system and attitude control system would likely reduce performance, we believe that the comparison highlights the advantages of using Meta-RL to optimize an adaptive GN\&C policy.  We also test the ability of the optimized policy to generalize to novel initial conditions and target  maneuvers. Finally, we test the policy with a substantially modified inertia tensor and a fuel slosh model to demonstrate the ability of Meta-RL to adapt to differences between the simulator used for optimization and the actual environment experienced during deployment. Mapping observations to actions requires only four small matrix multiplications, which takes less than 1 ms on a 2.3 Ghz CPU and requires around 64KB of memory. In this work we use a 25 Hz guidance frequency, and we expect that the policy would easily run on modern flight computers.

\section{Problem Formulation}

\subsection{Missile Configuration}\label{missile_config}

The missile is modeled as a cylinder of height  $h=1 \text{ m}$ and radius $r=0.25\text{ m}$ about the missile's body frame x-axis  with a wet and dry mass $m$ of 50 kg and 25 kg, respectively, and inertia tensor as given in Equation \ref{eq:inertia_tensor}. 

\begin{equation}
    \label{eq:inertia_tensor}
    {\bf J}=m\begin{bmatrix} r^2 / 2 & 0 & 0 \\ 0 & (3r^2  + h^2)/12 & 0 \\ 0 & 0 & (3r^2+h^2)/12\end{bmatrix}
\end{equation}

Four divert thrusters  thrusters and 12 attitude control thrusters are positioned as shown in Table \ref{tab:thrusters}. The attitude control thrusters operate in pairs, i.e., firing thrusters 4 and 5 cause a clockwise torque around the  missile's x-axis, whereas  firing thrusters 6 and 7 together create a counter-clockwise torque around the x-axis. Each divert thruster creates 5000 N of force, whereas the attitude control thrusters each create 125 N of force. The  thrusters can be switched on or off at the guidance frequency of 25 Hz. With a 5\% shift in the missile's center of mass (caused by fuel consumption), the torques caused by a divert thrust can be exactly cancelled by firing the appropriate attitude control thrusters. For center of mass variation less than 5\%, the attitude control thrusters will overcompensate for the torque induced by the divert thrust. The nominal (wet mass) missile center of mass is assumed to be [0,0,0] in body frame coordinates, and we define center of mass variation as a percentage of the missile dimensions, i.e., a 5\% variation would offset the center of mass by +/- 2.5 cm (0.05 * $h$/2) in the body frame $x$ direction and 1.25 cm (0.05 * $r$) in the $y$ and $z$ body frame directions. The instantaneous center of mass is as shown in Equation \ref{eq:COM}, where $\mathbf{r}_{\mathrm{com}}(t)$ is the instantaneous center of mass at time $t$, $\mathbf{r}_{\mathrm{com}}(t_o) \in \mathbb{R}^3$ is chosen from a uniform distribution at the start of an episode within the range given in Table \ref{tab:ic}, $f_{used}$ is the fuel used up to time $t$, and $f_{max}$ is the amount of fuel at the start of the engagement (25kg). 

\begin{equation}
	\label{eq:COM}
	\mathbf{r}_{\mathrm{com}}(t) = (\mathbf{r}_{\mathrm{com}}(t_o)) (f_{used}) / (f_{max})
\end{equation}

During policy testing, we also define an alternate center of mass variation model where the center of mass offset changes randomly at each time step of the episode. This is meant to very roughly account for fuel slosh.  While not an accurate model of fuel slosh, it creates a more difficult parasitic effect as compared to Equation \ref{eq:COM}.

\begin{table}[h]
	\fontsize{10}{10}\selectfont
    \caption{Body Frame Thruster Locations.}
   \label{tab:thrusters}
        \centering 
   \begin{tabular}{c r  r  r  r  r  r  r } 
      \hline
      & \multicolumn{3}{c}{Direction Vector} & \multicolumn{3}{c}{Location} & \multicolumn{1}{c}{Rotation}  \\
       \hline
      Thruster & x  & y  & z  &  x (m) & y (m) & z (m) & Axis\\
      \hline
      1 & 0.00 & -1.00 & 0.00 & 0.00 & -0.25 & 0.00  & N/A  \\
      2 & 0.00 & 1.00 & 0.00 & 0.00 & 0.25 & 0.00 & N/A \\
      3 & 0.00 & 0.00 & 1.00 & 0.00 & 0.00 & 0.25 & N/A\\
      4 & 0.00 & 0.00 & -1.00 & 0.00 & 0.00 & -0.25 & N/A \\
      
      5 & 0.00 & 0.00 &  1.00 & 0.00 & -r & 0.00 &  x  \\
      6 & 0.00 & 0.00 & -1.00 & 0.00 &  r & 0.00 &  x  \\

      7 & 0.00 & -1.00 &  0.00 & 0.00 & 0.00&  r &  x   \\   
      8 & 0.00 &  1.00 &  0.00 & 0.00 & 0.00& -r &  x    \\  

      9 & 0.00 &  0.00 & -1.00 &  h/2 & 0.00 & -r &  y \\
     10 & 0.00 &  0.00 &  1.00 & -h/2 & 0.00 &  r &  y \\

     11 & 0.00 & 0.00 &  1.00 &  h/2 & 0.00 &  r &  y     \\
     12 & 0.00 & 0.00 & -1.00 & -h/2 & 0.00 & -r &  y    \\  

     13 & 0.00 & -1.00 &  0.00 &  h/2 & -r & 0.00  &  z    \\  
     14 & 0.00 &  1.00 &  0.00 & -h/2 &  r & 0.00  &  z   \\  

     15 & 0.00 &  1.00 &  0.00 & h/2 &  r & 0.00  &  z    \\ 
     16 & 0.00 & -1.00 &  0.00 & -h/2 & -r & 0.00 &  z    \\ 

   \end{tabular}
\end{table}

In this work we assume the missile is equipped with a strapdown seeker with a 90 degree field of view (FOV). This allows the GN\&C system to keep the attitude stabilized such that the seeker boresight axis (which in our case is aligned with the missile centerline) points at where the target would be at the time of intercept in the absence of any heading or attitude error, i.e., the attitude at the end of a perfectly executed mid-flight phase, which also happens to point at the most likely location of the actual target.  This is a different arrangement from deployed exoatmospheric interceptors, where the missile attempts to keep its attitude such that the centerline points at the current target position. The rationale for this latter approach is to accommodate seekers with a limited FOV. However, modern infrared focal point arrays with millions of elements allow larger FOV. To illustrate, a wavelength of 12 um and 10cm aperture give an instantaneous FOV per pixel of $1.22 * 12\times10^{-6} / 0.1 = 146.4 \text{ ur}$, and a 100 mega-pixel array would then achieve a 84 degree field of view. Moreover, recent advances in curved focal plane arrays will allow large fields of view with fewer pixels. Finally, there is the possibility of combining multiple focal plane arrays with limited fields of view to extract the information that would be available with a single high FOV sensor (as with DARPA's ARGUS sensor). 

There are many advantages to our approach. First, if the lead angle $L$ in Figure \ref{fig:HE} (Section \ref{engagement}) is large, keeping the missile centerline pointed at the target makes the divert thrusts less orthogonal to to the missile's velocity vector, and therefore less effective. Second, at the initiation of the end game, the GN\&C system must discriminate the target from a background of clutter and potential decoys. If the missile centerline is rotated towards the first likely target, which might be a decoy, the actual target might then fall outside of the seeker's field of view.  But if the missile centerline is kept aligned with the expected intercept point, the target discrimination system can decide later in the endgame that there there is a more promising object to track. To be clear, we are not addressing target discrimination in this work. Another problem with keeping the missile pointed at the target is that the missile must rotate during the engagement to track the target. The resulting rotation in conjunction with scale factor errors will give a false indication of change in the stabilized seeker angles resulting in a parasitic attitude loop (see section \ref{PAL}). Finally, if the  target executes a spiral maneuver, the missile will need to constantly adjust its attitude to remain pointed at the target, wasting fuel, and potentially decreasing accuracy. In general, a larger FOV simplifies target acquisition and discrimination, and since endoatmospheric missiles with IR strapdown seekers cannot keep the seeker boresight pointed at the target, we can expect that in  the near future, focal plane arrays will be available with a 90 degree or greater FOV.

\subsection{Seeker Model}\label{seeker}

Given ground truth missile and target positions $\mathbf{r}_{\mathrm{M}}$ and $\mathbf{r}_{\mathrm{T}}$ we can define the relative position $\mathbf{r}_{\mathrm{TM}} = \mathbf{r}_{\mathrm{T}} - \mathbf{r}_{\mathrm{M}}$, and we denote the relative inertial frame line of sight unit vector as $\hat{\mathbf{r}}_{\mathrm{TM}}^N = \mathbf{r}_{\mathrm{TM}} / \|\mathbf{r}_{\mathrm{TM}}\|$. Defining $\mathbf{C}_\mathrm{BN}(\mathbf{q})$ as the direction cosine matrix (DCM) mapping from the inertial frame to the body frame given the missile's current attitude $\bf q$, the body frame line of sight unit vector is then computed as $\hat{\mathbf{r}}_{\mathrm{TM}}^B = \mathbf{C}_\mathrm{BN}(\mathbf{q})\hat{\mathbf{r}}_{\mathrm{TM}}^N$.  We can then compute the ground truth body frame target elevation and azimuth angles $\theta_{v_{\mathrm{GT}}}^B$ and $\theta_{u_{\mathrm{GT}}}^B$ as the orthogonal projection of $\hat{\mathbf{r}}_{\mathrm{TM}}^B$ onto the body frame unit vectors  $\hat{\mathbf{u}}=\begin{bmatrix} 0 & 1 & 0 \end{bmatrix}$, $\hat{\mathbf{v}}=\begin{bmatrix} 0 & 0 & 1\end{bmatrix}$, as shown in Equations \ref{eq:los1} and \ref{eq:los2}. Note that in the actual missile hardware, $\theta_{u_{\mathrm{GT}}}^B$ and $\theta_{v_{\mathrm{GT}}}^B$ would be estimated in the target discrimination system (See Figure \ref{fig:System}) based off of the pixel location of the assumed target on the focal plane array. 

\begin{subequations}
\begin{align}
\theta_{u_{\mathrm{GT}}}^B &= \mathrm{arcsin}(\hat{\mathbf{r}}_{\mathrm{TM}}^B \cdot \hat{\mathbf{u}})\label{eq:los1}\\
\theta_{v_{\mathrm{GT}}}^B &= \mathrm{arcsin}(\hat{\mathbf{r}}_{\mathrm{TM}}^B \cdot \hat{\mathbf{v}})\label{eq:los2}
\end{align}
\end{subequations}

We use a noise model where the missile ground truth rotational velocity $\boldsymbol{\omega}$ and the ground truth body-frame seeker angles $\theta_{u_{\mathrm{GT}}}^B$ and $\theta_{v_{\mathrm{GT}}}^B$ are corrupted by a scale factor error (as described in \cite{willman1988effects}) and Gaussian noise, as shown below in Equations \ref{eq:w_noise} through \ref{eq:angle_noise2}, where $\mathcal{N}(0,\sigma,n)$ is an $n$ dimensional zero mean Gaussian random variable with diagonal covariance matrix with diagonal elements set to $\sigma^2$, $e_{\omega}$ is the rotational velocity scale factor error, and $e_{\theta}$ is the seeker angle scale factor error.  

The simulator provides $\hat{\boldsymbol{\omega}}$, $\theta_{u}^B$ and $\theta_{v}^B$ as inputs to the GN\&C system. 

\begin{subequations}
\begin{align}
	\hat{\boldsymbol{\omega}} &= \boldsymbol{\omega}_{GT} (1+e_{\omega}) + \mathcal{N}(0,\sigma_{\omega},3)\label{eq:w_noise}\\
	\theta_{u}^B &= \theta_{u_{GT}}^B (1+e_{\theta}) + \mathcal{N}(0,\sigma_{\theta},1)\label{eq:angle_noise1} \\
	\theta_{v}^B &= \theta_{v_{GT}}^B (1+e_{\theta}) + \mathcal{N}(0,\sigma_{\theta},1)\label{eq:angle_noise2}
\end{align}
\end{subequations}

Since $\theta_{u}^B$ and $\theta_{v}^B$ are in the missile body frame, which in general can be rotating, the GN\&C system must rotate them back to an inertial reference frame so that missile rotations are not confused with target maneuvers. Moreover, a strapdown seeker cannot be mechanically stabilized, as it is fixed in the missile body frame. Instead, we use the estimated change in attitude since the initiation of the homing phase ($\bf dq$) to rotate $\theta_v^B$ and $\theta_u^B$  back to the missile's attitude at the start of the homing phase, which defines an inertial reference frame that we will refer to as $N'$, where in general $N'$ is rotated from $N$. Specifically, we start by computing the reconstructed line of sight direction vector $\hat{\boldsymbol{\lambda}}_r^B$ as shown in Equations \ref{eq:seeker1} through \ref{eq:seeker4}.

\begin{subequations}
\begin{align}
y &= \sin(\theta_u^B) \label{eq:seeker1}\\
z &= \sin(\theta_v^B) \label{eq:seeker2}\\
x &= \sqrt{1-z^2-y^2} \label{eq:seeker3}\\
\hat{\boldsymbol{\lambda}}_r^B &= [x,y,z] \label{eq:seeker4}\\
\end{align}
\end{subequations}

Further, we define  $\mathbf{C}_\mathrm{BN'}(\mathbf{dq})$ as the DCM mapping from the inertial frame $N'$ to the body frame given $\bf dq$. We can now define the stabilized seeker angles $\theta_u^S$ and $\theta_v^S$, and compute them as shown in Equations \ref{eq:seeker5} through \ref{eq:seeker7}. 

\begin{subequations}
\begin{align}
\hat{\boldsymbol{\lambda}}^S &= \mathbf{C}_\mathrm{BN'}(\mathbf{dq})^T \hat{\boldsymbol{\lambda}}_r^B \label{eq:seeker5}\\
\theta_{u}^S &= \mathrm{arcsin}(\boldsymbol{\hat\lambda}^\mathrm{S} \cdot \hat{\mathbf{u}})\label{eq:seeker6}\\
\theta_{v}^S &= \mathrm{arcsin}(\boldsymbol{\hat\lambda}^\mathrm{S} \cdot \hat{\mathbf{v}})\label{eq:seeker7}
\end{align}
\end{subequations}

We then pass $\theta_{u}^S$ and $\theta_{u}^S$ through a first order lag filter with time constant $\tau_{\theta}$ as shown in Equations \ref{eq:theta_u_lag} and \ref{eq:theta_v_lag}. The lag filter attenuates the Gaussian noise $\sigma_{\theta}$, allowing a better estimate of the rate of change of $\theta_{u}^S$ and $\theta_{v}^S$. In addition, the lag filter damps the response to the parasitic attitude loop described in Section \ref{PAL}.

\begin{subequations}
\begin{align}
	\dot{\theta}_{u} &= (\theta_{u}^S - \theta_{u}) / \tau_{\theta}\label{eq:theta_u_lag}\\
	\dot{\theta}_{v} &= (\theta_{v}^S - \theta_{v}) / \tau_{\theta}\label{eq:theta_v_lag}
\end{align}
\end{subequations}

We define the estimated seeker angles as $\hat{\theta}_u \equiv {\theta}_u$ and $\hat{\theta}_v\equiv {\theta}_v$, where $\theta_u$ and $\theta_v$ are obtained by integrating Equations \ref{eq:theta_u_lag} and \ref{eq:theta_v_lag}. We compute the estimated seeker angles rates of change as shown in Equations \ref{eq:theta_u_dot} and \ref{eq:theta_v_dot}, where $dt$ is the guidance period (40ms).

\begin{subequations}
\begin{align}
	\hat{\dot{\theta}}_u = \frac{(\hat{\theta}_u(t) - \hat{\theta}_u(t-1))}{dt}\label{eq:theta_u_dot}\\
    \hat{\dot{\theta}}_v = \frac{(\hat{\theta}_v(t) - \hat{\theta}_v(t-1))}{dt}\label{eq:theta_v_dot}
\end{align}
\end{subequations}

Since we assume the change in attitude cannot be directly measured, we must integrate $\hat{\boldsymbol{\omega}}$ to obtain an estimate of $\bf{dq}$ as shown in Equation \ref{eq:dqi}.  In our simulation model, we approximate the integration using fourth order Runge-Kutta integration with a 20 ms timestep.  Assuming that our rate gyros can provide output at a rate of at least 75 Hz, this could be implemented on the flight computer as our guidance frequency is 25 Hz. The Meta-RL policy described in Section \ref{GLD} will map $\hat{\theta}_u$ and $\hat{\theta}_v$,  $\hat{\dot{\theta}}_u$,  $\hat{\dot{\theta}}_v$, $\hat{\mathbf{dq}}$, and $\hat{\boldsymbol{\omega}}$ directly to  divert thrust commands. 

\begin{equation}
	\label{eq:dqi}
	\hat{\mathbf{dq}}(t) = \int_0^t \hat{\boldsymbol{\omega}}(t) dt
\end{equation}

\subsection{Parasitic Attitude Loop}\label{PAL}

Our attitude stabilization scheme is not immune to the effects of scale factor errors and Gaussian noise, where missile rotation can lead to a false indication of changes in the stabilized seeker angles, with the missile rotation being caused by either divert thrusts with a changing center of mass or the attitude control thrust actions that attempt to correct for this. The guidance system will attempt to nullify these (false) changes, and the resulting divert thrusts will result in further rotation due to changes in the missile's center of mass from fuel consumption.  Attitude control thrusters will attempt to correct for these rotations, but the correction will not be perfect due to the pulsed thrusters, resulting in under or over correction of the divert thrust induced rotations. These rotations then lead to further false indications of change in the stabilized seeker angles. 

Looking at this in more detail, we first consider the rotational velocity scale factor error $e_{\omega}$ for the case where in an inertial (i.e.,  perfectly stabilized) reference frame, the line of sight to target is constant, but the missile is rotating. In this scenario, perfectly stabilized seeker angles would be constant. But in order to compensate for the missile rotations, perfect stabilization requires an error free estimate of $\mathbf{dq}$. Although the integrator will have a filtering effect on $e_{\omega}$ when a sequence of $\hat{\boldsymbol{\omega}}$ measurements has low temporal correlation, let's look at the worst case scenario where $\hat{\boldsymbol{\omega}}$ is constant. In this case our estimate of $\mathbf{dq}$ will be $\hat{\mathbf{dq}} = \mathbf{dq}(1+e_{\omega})$, and the rotation used to stabilize our seeker angles will be inaccurate. This will result in a changing $\hat{\theta}_u$ and $\hat{\theta}_v$, giving a non-zero value for $\hat{\dot{\theta}}_u$ and $\hat{\dot{\theta}}_v$ that the guidance policy will attempt to nullify. 

Seeker angle scale factor errors $e_{\theta}$ will also cause false indications of change in the stabilized seeker angles when the body-frame line of sight to target is reconstructed as shown in Equations \ref{eq:seeker1} through \ref{eq:seeker3}. By substituting Equations \ref{eq:angle_noise1} and \ref{eq:angle_noise2} into Equations \ref{eq:seeker1} through \ref{eq:seeker3}, we see that $e_{\theta}$ will induce a rotation in $\hat{\boldsymbol{\lambda}}_r^B$, which in turn will result in a non-zero rotation of the stabilized seeker angles, even with a perfect estimation of $\mathbf{dq}$.

It is worth noting that even with a perfect estimate of $\mathbf{dq}$ and a perfect reconstruction of $\hat{\boldsymbol{\lambda}}_r^B$, $e_{\theta}$ will still cause false indications of change in the stabilized seeker angles.  Substituting Equations \ref{eq:angle_noise1} and \ref{eq:angle_noise2} into Equations \ref{eq:theta_u_dot} and \ref{eq:theta_v_dot}, we see in Equation \ref{eq:radome} that perfect stabilization will set the first term to zero, but not the second, so we get a false rotation proportional to $e_{\theta}$.

\begin{equation}
	\label{eq:radome}
	\hat{\dot{\theta}}_u = \frac{\left(\theta_u(t) - \theta_u(t-1)\right) + \left(\theta_u(t)e_{\theta} - \theta_u(t)e_{\theta}\right)}{dt}\\
\end{equation}

\subsection{Engagement Scenario}\label{engagement}
\begin{figure}[h]
\begin{center}
\includegraphics[width=.6\linewidth]{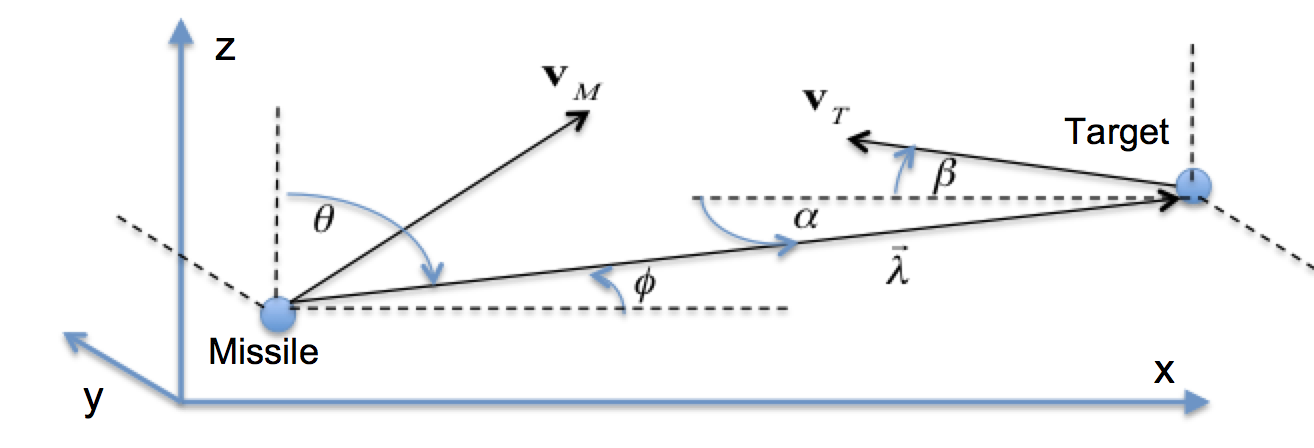}
\caption{Engagement}
\label{fig:engagement}
\end{center}
\end{figure}

In this work we use a simplified engagement scenario. Instead of modeling the missile and target trajectories using Lambert guidance to determine BRV trajectory and Kepler guidance to determine the missile trajectory, the engagement is modeled as a simple skewed head-on engagement, with the collision triangle modified to account for the gravitational field. Referring to Fig. \ref{fig:engagement}\footnote{In this figure, the illustrated vectors are not in general within the y-z plane}, where the missile velocity vector, target velocity vector, and relative range vector are given as $\mathbf{v}_\mathrm{M}$, $\mathbf{v}_\mathrm{T}$, and $\mathbf{r_\mathrm{TM}}$, we can define the target's initial position $\mathbf{r}_\mathrm{T}$ in spherical coordinates in a missile centered reference frame, in terms of the range from missile to target $\|\bf r_\mathrm{TM}\|$ and angles $\theta$ and $\phi$ as given in Equations \eqref{eq:eng1}, \eqref{eq:eng2}, and \eqref{eq:eng3}.

\begin{subequations}
\begin{align}
r_{\mathrm{T}_x} &= \|\bf r_\mathrm{TM}\|\mathrm{sin}(\theta)\mathrm{cos}(\phi)\label{eq:eng1}\\
r_{\mathrm{T}_y} &= \|\bf r_\mathrm{TM}\|\mathrm{sin}(\theta)\mathrm{sin}(\phi)\label{eq:eng2}\\
r_{\mathrm{T}_z} &= \|\bf r_\mathrm{TM}\|\mathrm{cos}(\theta)\label{eq:eng3}
\end{align}
\end{subequations}

Further, we can represent the target's initial velocity vector $\mathbf{v}_\mathrm{T}$ in terms of the magnitude of the target velocity $\|\mathbf{v}_\mathrm{T}\|$ and angles $\alpha$ and $\beta$ as shown in Equations \eqref{eq:he1}, \eqref{eq:he2}, and \eqref{eq:he3}.

\begin{subequations}
\begin{align}
v_{\mathrm{T}_x} &= -\|\mathbf{v}_\mathrm{T}\|\mathrm{cos}(\beta)\mathrm{cos}(\alpha)\label{eq:he1}\\
v_{\mathrm{T}_y} &= -\|\mathbf{v}_\mathrm{T}\|\mathrm{cos}(\beta)\mathrm{sin}(\alpha)\label{eq:he2}\\
v_{\mathrm{T}_z} &= \|\mathbf{v}_\mathrm{T}\|\mathrm{sin}(\beta)\label{eq:he3}
\end{align}
\end{subequations}

\begin{figure}[h]
\begin{center}
\includegraphics[width=.6\linewidth]{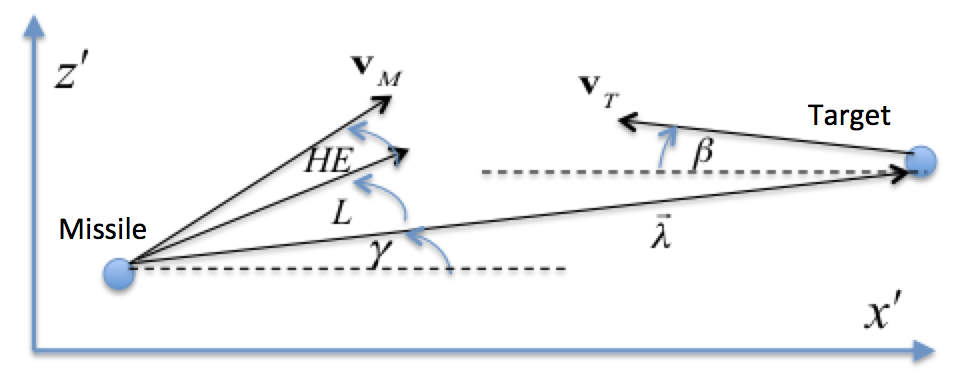}
\caption{Planar Heading Error}
\label{fig:HE}
\end{center}
\end{figure}

A collision triangle can be defined in a plane that is not in general aligned with the coordinate frame shown in Fig. \ref{fig:engagement}, and is illustrated in Fig. \ref{fig:HE}.  Here we define the required lead angle $L$ for the missile's velocity vector $\mathbf{v}_m$ as the angle that will put the missile on a collision triangle with the target in terms of the target velocity $\mathbf{v}_\mathrm{T}$, time of flight $t_f=\|\mathbf{r}_{\mathrm{TM}}\| / v_c$, the gravitational acceleration acting on the target $\mathbf{g}_{\mathrm{T}}$, line-of-sight angle $\gamma$, and the magnitude of the missile velocity as shown in Equations~\eqref{eq:lead1} through \eqref{eq:lead3}. Closing velocity is computed as $v_c = -\mathbf{r}_{\mathrm{TM}} \cdot \mathbf{v}_{\mathrm{TM}} / \|\mathbf{r}_{\mathrm{TM}}\|$.  Since $t_f$ is not known until we compute the required lead angle and ultimately $\mathbf{v}_m$, we use a two step approach, where we first get the required lead angle neglecting gravity, use this to compute $\mathbf{v}_m$ and $t_f$, and then recompute the required lead angle and $\mathbf{v}_m$ using this estimated time of flight and $\mathbf{g}_{\mathrm{T}}$ (see Equations \ref{eq:g1} through \ref{eq:g6}). Although this does not create a perfect collision triangle accounting for $\bf{g}_T$, it gets close (less than 5m open loop miss distance with no heading error or target acceleration).

\begin{subequations}
\begin{align}
    L &= \arcsin\left(\frac{(\|\mathbf{v}_\mathrm{T}\|+t_f\mathbf{g}_\mathrm{T})\sin(\beta+\gamma)}{\|\mathbf{v}_\mathrm{M}\|}\right)\label{eq:lead1}\\
    v_{m_y} &= \|\mathbf{v}_{\mathrm{M}}\|\cos(L + \gamma)\label{eq:lead2}\\
    v_{m_z} &= \|\mathbf{v}_{\mathrm{M}}\|\sin(L + \gamma)\label{eq:lead3}
\end{align}
\end{subequations}

This formulation is easily extended to a three dimensional engagement using the following approach:

\begin{enumerate}
\item define a plane normal as $\mathbf{\hat{v}}_t \times \boldsymbol{\hat{\lambda}}$
\item rotate $\mathbf{v}_{\mathrm{T}}$ and $\boldsymbol{\hat{\lambda}}$ onto the plane
\item calculate the required planar missile velocity (Equations \eqref{eq:lead1} through \eqref{eq:lead3})
\item rotating this velocity back into the original reference frame
\end{enumerate}

Thus in $\mathbb{R}^3$ we define a heading error (HE) as the the angle between the missile's initial velocity vector and the velocity vector associated with the lead angle required to put the missile on a collision heading with the target, accounting for the gravitational field. We also define the initial attitude error as the angle between the missile's velocity vector and the body frame x-axis at the start of the engagement.

\begin{figure}[h]
\begin{center}
\includegraphics[width=1.0\linewidth]{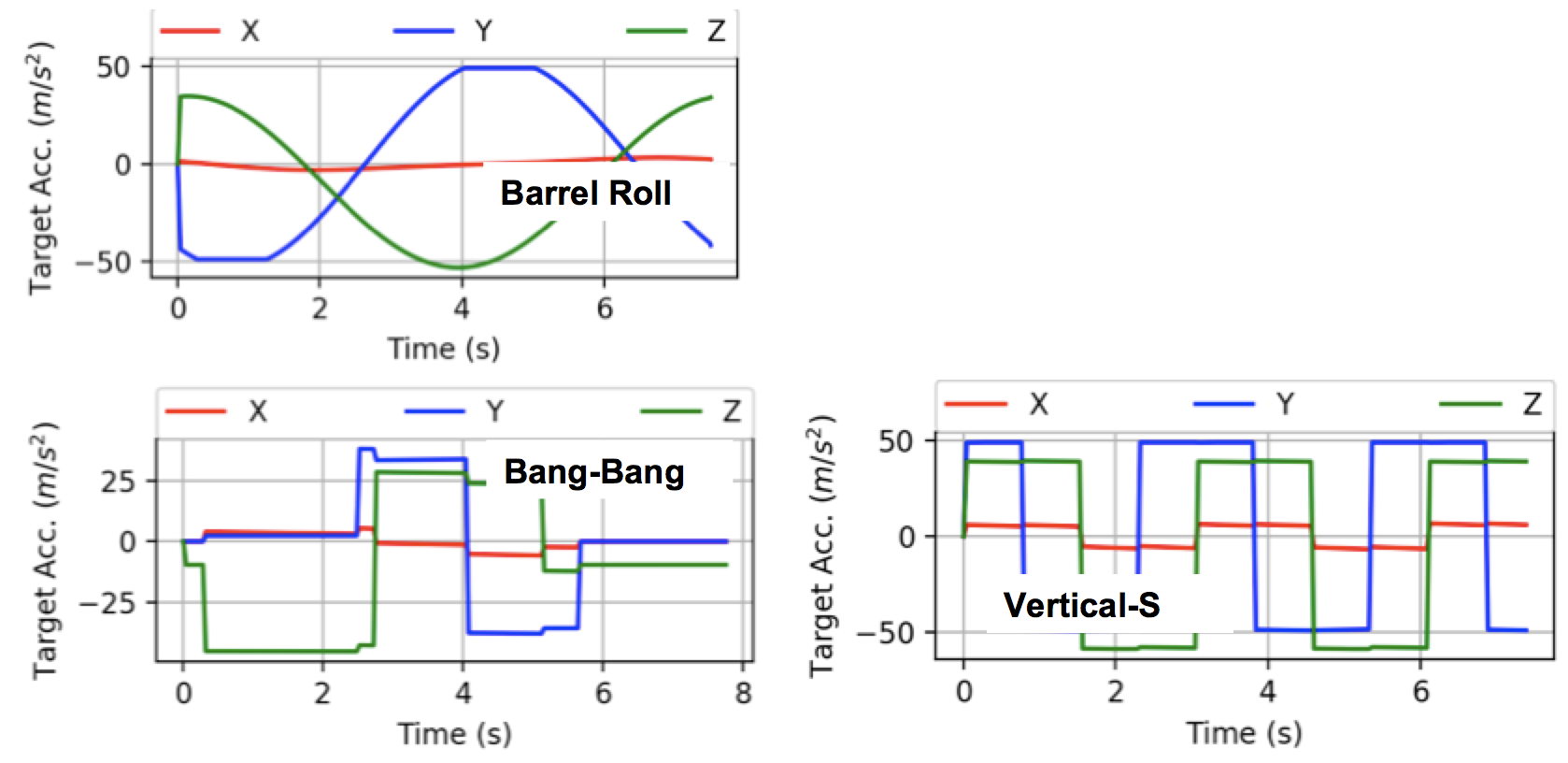}
\caption{Sample Target Maneuvers}
\label{fig:TM}
\end{center}
\end{figure}

During optimization we randomly choose between a target bang-bang and vertical-S target maneuver with equal probability, with the acceleration applied orthogonal to the target's velocity vector. The maneuvers have varying acceleration levels up to a maximum of $\mathrm{5*9.81\ m/s}^2$, and with random start time, duration, and switching time. For testing, we also introduce a barrel roll maneuver. Sample target maneuvers are shown Fig. \ref{fig:TM}, note that in some cases the maneuver period is considerably shorter or longer.

We can now list the range of engagement scenario parameters in Table \ref{tab:ic}.  During optimization and testing, these parameters are drawn uniformly between their minimum and maximum values.  The generation of heading error is handled as follows. We first calculate the optimal missile velocity vector that puts the missile on a collision triangle with the target as described previously. We then uniformly select a heading error $HE$ between the bounds given in Table \ref{tab:ic}, and randomly perturb the missile's velocity vector's direction such that $\arccos({\mathbf{v}_{\mathrm{M}} \cdot \mathbf{v}_{{\mathrm{M}}_p}}) < HE  $, where $\mathbf{v}_{\mathrm{M}_p}$ is the perturbed missile velocity vector. Similarly, we define an ideal initial attitude as having the missile's x-axis aligned with its velocity vector, and use a similar method to randomly perturb this by the attitude error such that angle between the missile's ideal and actual velocity vector is less than or equal to the attitude error. 

\begin{table}[h]
	\fontsize{10}{10}\selectfont
    \caption{Initial Conditions}
   \label{tab:ic}
        \centering 
   \begin{tabular}{l  r  r } 
      \hline
      Parameter & min & max \\
      \hline
      Range $\|\mathbf{r}_\mathrm{TM}\|$ (km) & 50 & 55\\
      Missile Velocity Magnitude (m/s) & 3000 & 3000 \\
      Target Position angle $\theta$ (degrees) & 80 & 100 \\
      Target Position angle $\phi$ (degrees) & -10 & 10 \\
      Target Velocity Magnitude (m/s) & 4000 & 4000 \\
      Target Velocity angle $\beta$ (degrees) & -10 & 10 \\
      Target Velocity angle $\alpha$ (degrees) & -10 & 10 \\
      Heading Error (degrees) & 0 & 5 \\
      Attitude Error (degrees) & 0 & 5 \\
      Target Maximum Acceleration  $\mathrm{m/s^2}$ & 0 & 5*9.81\\
      Target Bang-Bang duration  (s) & 1 & 4 \\
      Target Bang-Bang initiation time (s) & 0 & 6 \\
      Target Barrel Roll / Vertical-S Period (s) & 1 & 5 \\
      Target Barrel Roll / Vertical-S Offset (s) & 1 & 5 \\
      \hline
      Center of Mass Variation $r_{\mathrm{com}}$  (\%) & -2.5 & 2.5\\
      Seeker Angle Scale Factor Error $e_{\theta}$  & $-1\times10^{-3}$ & $1\times10^{-3}$\\
      Seeker Angle Gaussian Noise $\sigma_{\theta}$ (rad)  & $1\times10^{-3}$ & $1\times10^{-3}$\\
      Rotational Velocity Scale Factor Error $e_{\omega}$   & $-1\times10^{-3}$ & $1\times10^{-3}$\\
      Rotational Velocity Gaussian Noise  $\sigma_{\omega}$ (rad) & $1\times10^{-3}$ & $1\times10^{-3}$\\
      Thruster Ignition Time Constant $\tau_{\mathrm{u}}$ ms & 20 & 20\\
      Angle Filter Time Constant $\tau_{\theta}$ ms & 20 & 20\\
   \end{tabular}
\end{table}

\subsection{Equations of Motion}\label{EQOM}

The force $\mathbf{F}_{B}$ and torque $\mathbf{L}_{B}$ in the missile's body frame for a given commanded thrust depends on the placement of the thrusters in the missile structure. We can describe the placement of each thruster through a body-frame direction vector $\mathbf{d}$ and position vector $\mathbf{r}$, both in $\mathbb{R}^3$.  The direction vector is a unit vector giving the direction of the body frame force that results when the thruster is fired.  The position vector gives the body frame location with respect to the missile centroid,   where the force resulting from the thruster firing is applied for purposes of computing torque, and in general the center of mass ($\mathbf{r}_\mathrm{com}$) varies with time as fuel is consumed. For a missile with $k$ thrusters, the body frame force and torque associated with one or more  thrusters firing is then as shown in Equations \eqref{eq:Thruster_modela} and \eqref{eq:Thruster_modelb}, where $T_{\mathrm{com}}^{i}$ is the commanded thrust for thruster $i$, $\mathbf{d}^{(i)}$ the direction vector for thruster $i$, and $\mathbf{r}^{(i)}$ the position of thruster $i$. The total body frame force and torque are calculated by summing the individual forces and torques.

\begin{subequations}
\begin{align}
    \tilde{\mathbf{F}}_{B}&=\sum_{i=1}^{k} \mathbf{d}^{(i)} \mathbf{T}_{\mathrm{com}}^{(i)} \label{eq:Thruster_modela}\\
	\tilde{\mathbf{L}}_{B}&=\sum_{i=1}^{k}(\mathbf{r}^{(i)}-\mathbf{r}_\mathrm{com})\times\tilde{\mathbf{F}}_{B}^{(i)}\label{eq:Thruster_modelb}
\end{align}
\end{subequations}

The force and torque are then passed through a first order lag simulated by integrating Equations~\ref{eq:T_lag1} and \ref{eq:T_lag2} , where $\tau_{\mathrm{u}}$ is the time constant of the first order lag. This models the thruster ignition lag.

\begin{subequations}
\begin{align}
	\mathbf{\dot{F}}_{B}=(\tilde{\mathbf{F}}_{B} - \mathbf{F}_{B}) / \tau_{\mathrm{u}}\label{eq:T_lag1}\\
	\mathbf{\dot{L}}_{B}=(\tilde{\mathbf{L}}_{B} - \mathbf{L}_{B}) / \tau_{\mathrm{u}}\label{eq:T_lag2}
\end{align}
\end{subequations}

The dynamics model uses the missile's current attitude $\mathbf{q}$ to convert the body frame thrust vector to the inertial frame as shown in in Equation \eqref{eq:BtoN} where $\mathbf{C}_\mathrm{BN}(\mathbf{q})$ is the direction cosine matrix mapping the inertial frame to body frame obtained from the current attitude parameter $\mathbf{q}$.

\begin{equation}
	\label{eq:BtoN}
	\mathbf{F}_{N}=\left[\mathbf{C}_\mathrm{BN}(\mathbf{q})\right]^{T}\mathbf{F}_{B}
\end{equation}

The rotational velocities $\bm{\omega}_{B}$ are then obtained by integrating the Euler rotational equations of motion, as shown in Equation \eqref{eq:EulerRot}, where $\mathbf{L}_{B}$ is the body frame torque as given in Equation \eqref{eq:Thruster_modelb}, and $\mathbf{J}$ is the lander's inertia tensor. Note we have included a term that models a rotation induced by a changing inertia tensor, which in general is time varying as the missile consumes fuel.  Specifically, the inertia tensor is recalculated at each time step to account for fuel consumption, but we do not modify the inertia tensor to account for changes in the missile's center of mass.

\begin{equation}
	\label{eq:EulerRot}
	\mathbf{J}{\dot{\bm{\omega}}_{B}}=-\Tilde{\bm{\omega}}_{B}\mathbf{J}\bm{\omega}_{B}-\dot{\mathbf{J}}\bm{\omega}+\mathbf{L}_{B}
\end{equation}

The lander's attitude is then updated by integrating the differential kinematic equations shown in Equation \eqref{eq:diffeqom}, where the lander's attitude is parameterized using the quaternion representation and $\bm{\omega}_{i}$ denotes the $i^{th}$ component of the rotational velocity vector $\bm{\omega}_{B}$. 

\begin{equation}
    \label{eq:diffeqom}
    \begin{bmatrix} \dot{q_{0}} \\ \dot{q_{1}} \\ \dot{q_{2}} \\ \dot{q_{3}}\end{bmatrix} = \frac{1}{2}\begin{bmatrix} q_{0} & -q_{1} & -q_{2} & -q_{3}\\ q_{1} & q_{0} & -q_{3} & q_{2}\\ q_{2} & q_{3} & q_{0} & -q_{1} \\ q_{3} & -q_{2} & q_{1} & q_{0} \end{bmatrix} \begin{bmatrix} 0 \\ \omega_{0} \\ \omega_{1} \\ \omega_{2} \end{bmatrix}
\end{equation}

The missile's translational motion is modeled as shown in \ref{eq:EQOMa} through \ref{eq:EQOMc}. 

\begin{subequations}
\begin{align}
	{\Dot{\mathbf r}} &= {{\mathbf v}}\label{eq:EQOMa}\\
	{\Dot{\bf v}} &= \frac{{{\bf F}_{N}}}{m} + \bf{g}_M\label{eq:EQOMb}\\
	\Dot{m} &= -\frac{\sum_{i}^{k}\lVert{{\bf F}_{B}}^{(i)}\rVert}{I_\text{sp}g_\text{ref}} \label{eq:EQOMc}
\end{align}
\end{subequations}
Here  ${{\bf F}_{N}}^{(i)}$ is the inertial frame force as given in Equation~\eqref{eq:BtoN}, $k$ is the number of thrusters, $g_\text{ref}=9.81$ $\text{m}/\text{s}^{2}$,  $\mathbf{r}$ is the missile's position in the engagement reference frame. To calculate the gravitational acceleration acting on the missile $\bf{g}_M$, we translate the engagement reference frame to an earth centered reference frame as shown in Equations \ref{eq:g1} through \ref{eq:g6}, where $R$ is the Earth's radius, $\mu$ the Earth's gravitational parameter, $\theta$ is latitude, and $\phi$ is longitude.  For optimization, without any loss of generality, we model a polar intercept at an altitude of 50km.  To demonstrate that the policy is invariant to the the actual location of the intercept, we also test the optimized policy using an equatorial intercept at 1000km.

\begin{subequations}
\begin{align}
	x_E = R\sin(\theta)\cos(\phi)\label{eq:g1}\\
	y_E = R\sin(\theta)\sin(\phi)\label{eq:g2}\\
	z_E = R\cos(\theta)\label{eq:g3}\\
	\mathbf{r}_{\mathrm{E}} = [x_E,y_E,z_E]\label{eq:g4}\\
	\mathbf{r}'_{M} = \mathbf{r}_{\mathrm{E}} + \mathbf{r}_{\mathrm{M}}\label{eq:g5} \\
	\mathbf{g}_M = -\mu  \mathbf{r}'_{\mathrm{M}} / \|\mathbf{r}'_{\mathrm{M}}\|^3\label{eq:g6}
\end{align}
\end{subequations}

The target is modeled as shown in Equations~\eqref{eq:TEQOMa} and \eqref{eq:TEQOMb}, where $\mathbf{a}_{\mathrm{T}_\mathrm{com}}$ is the commanded acceleration for the target maneuver, and $\mathbf{g}_{\mathrm{T}}$ the gravitational acceleration acting on the target.

\begin{subequations}
\begin{align}
	{\Dot{\mathbf r}} &= {{\mathbf v}}\label{eq:TEQOMa}\\
	{\Dot{\mathbf v}} &= \mathbf{a}_{\mathrm{T}_\mathrm{com}} + \mathbf{g}_{\mathrm{T}}\label{eq:TEQOMb}
\end{align}
\end{subequations}

The equations of motion are updated using fourth order Runge-Kutta integration.  For ranges greater than 1000 m, a timestep of 20 ms is used, and for the final 1000 m of homing, a timestep of 0.067 ms is used in order to more accurately measure miss distance; this technique is borrowed from \cite{zarchan2012tactical:4}.  For the APN guidance law, this results in a 100\% hit rate (miss $<$ 50 cm) with no target maneuver, $\tau_{\theta}=\tau{u}=0$, and zero heading error at a guidance frequency of 25 Hz. We also checked that decreasing the integration step size did not improve performance for the augmented APN guidance law.

\section{IGN\&C System Optimization}\label{GLD}

\subsection{Reinforcement Learning Framework}\label{RL}

\begin{figure}[h]
\begin{center}
\includegraphics[width=.75\linewidth]{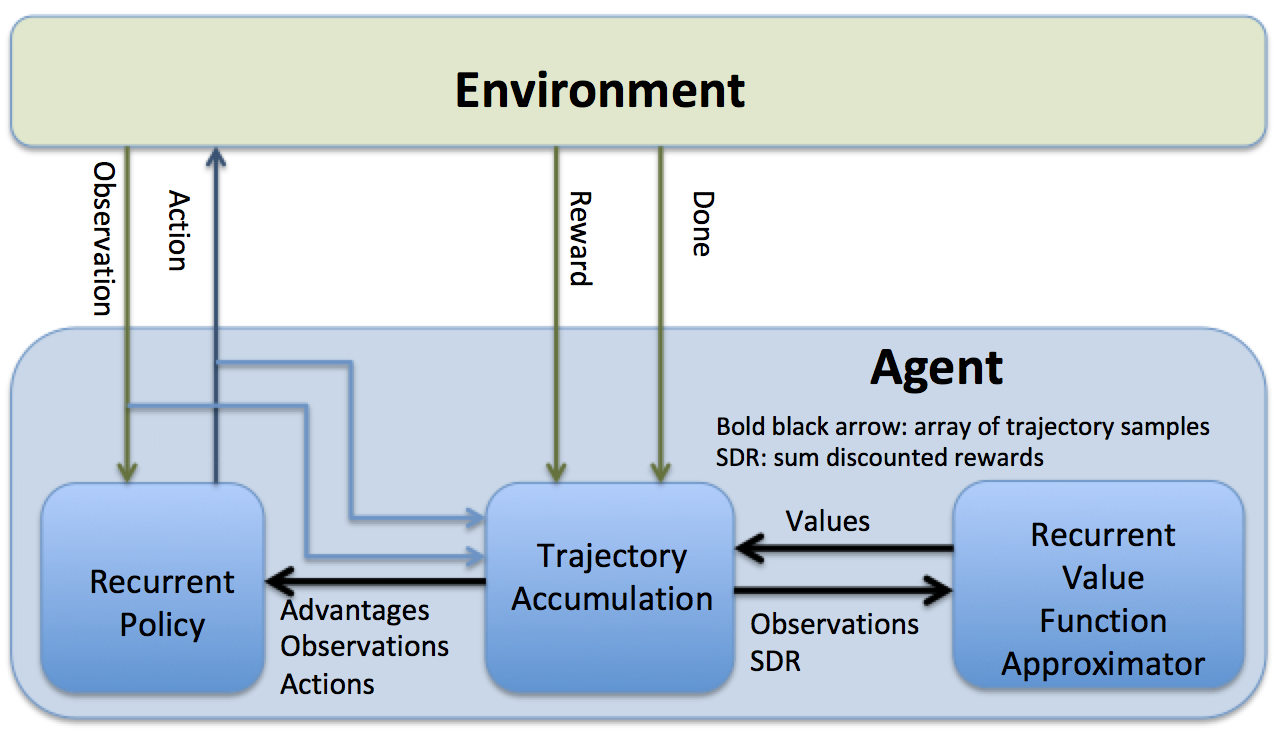}
\caption{Engagement}
\label{fig:Agent_env_detail}
\end{center}
\end{figure}

In the RL framework, an agent learns through episodic interaction with an environment how to successfully complete a task by learning a policy that maps observations to actions. The environment initializes an episode by randomly generating a ground truth state, mapping this state to an observation, and passing the observation to the agent.  The agent uses this observation to generate an action that is sent to the environment; the environment then uses the action and the current ground truth state to generate the next state and a scalar reward signal.  The reward and the observation corresponding to the next state are then passed to the agent. The process repeats until the environment terminates the episode, with the termination signaled to the agent via a done signal. In this application termination conditions include FOV violations (which would occur for both a successful or unsuccessful intercept) and any component of the missile's rotational velocity exceeding 12 rad/s. Trajectories collected over 30 episodes (referred to as rollouts) are collected during interaction between the agent and environment, and used to update the policy and value functions. The interface between agent and environment is depicted in Figure  \ref{fig:Agent_env_detail}, where the environment includes the modeled GN\&C system components, including the stabilization logic, seeker angle filter, and rotational velocity integrator.

A Markov Decision Process (MDP) is an abstraction of the environment, which in a continuous state and action space, can be represented by a state space $\mathcal{S}$, an action space $\mathcal{A}$, a state transition distribution $\mathcal{P}(\mathbf{x}_{t+1}|\mathbf{x}_t,\mathbf{u}_t)$, and a reward function $r=\mathcal{R}(\mathbf{x}_t,\mathbf{u}_t))$, where $\mathbf{x} \in \mathcal{S}$ and $\mathbf{u} \in \mathcal{A}$, and $r$ is a scalar reward signal. We can also define a partially observable MDP (POMDP), where the state $\mathbf{x}$ becomes a hidden state, generating an observation $\mathbf{o}$ using an observation function $\mathcal{O}(\mathbf{x})$ that maps states to observations. The POMDP formulation is useful when the observation consists of raw sensor outputs, as is the case in this work.  In the following, we will refer to both fully observable and partially observable environments as POMDPs, as an MDP can be considered a POMDP with an identity function mapping states to observations.

Reinforcement meta-learning differs from generic reinforcement learning in that the agent learns to quickly adapt to novel POMPDs by learning over a wide range of POMDPs. These POMDPs can include different environmental dynamics, target maneuvers, actuator failure scenarios, mass and inertia tensor variation, and varying amounts of sensor distortion. Learning within the RL meta-learning framework results in an agent that can quickly adapt to novel POMDPs, often with just a few steps of interaction with the environment. There are multiple approaches to implementing meta-RL.  In \cite{finn2017model}, the authors design the objective function to explicitly make the model parameters transfer well to new tasks. In \cite{mishra2017simple}, the authors demonstrate state of the art performance using temporal convolutions with soft attention. And  in \cite{frans2017meta}, the authors use a hierarchy of policies to achieve meta-RL. In this proposal, we use a different approach \cite{wang2016learning} using a recurrent policy and value function. Note that it is possible to train over a wide range of POMDPs using a non-meta RL algorithm. Although such an approach typically results in a robust policy, the policy cannot adapt in real time to novel environments. 

In this work, we  implement meta-RL using proximal policy optimization (PPO) \cite{schulman2017proximal} with both the policy and value function implementing recurrent layers in their networks.  To understand how recurrent layers result in an adaptive agent, consider that given some ground truth agent position, velocity, attitude, and rotational velocity $\mathbf{x}_{t}$, and action vector $\mathbf{u}_{t}$ output by the agent's policy, the next state $\mathbf{x}_{t+1}$ and observation $\mathbf{o}_{t+1}$ depends not only on $\mathbf{x}_{t}$ and $\mathbf{u}_{t}$, but also on the ground truth agent mass, inertia tensor, target maneuvers, and external forces acting on the agent. Consequently, during training, the hidden state of a network's recurrent network evolves differently depending on the observed sequence of observations from the environment and actions output by the policy. Specifically, the trained policy's hidden state captures unobserved (potentially time-varying) information such as external forces that are useful in minimizing the cost function. In contrast, a non-recurrent policy (which we will refer to as an MLP policy), which does not maintain a persistent hidden state vector, can only optimize using a set of current observations, actions, and advantages, and will tend to under-perform a recurrent policy on tasks with randomized dynamics \cite{gaudet2019adaptive}.  After training, although the recurrent policy's network weights are frozen, the hidden state will continue to evolve in response to a sequence of observations and actions, thus making the policy adaptive.  In contrast, an MLP policy's behavior is fixed by the network parameters at test time.

The PPO algorithm used in this work  is a  policy gradient algorithm which has demonstrated state-of-the-art performance for many RL benchmark problems. PPO approximates the TRPO optimization method \cite{schulman2015trust} by accounting for the policy adjustment constraint with a clipped objective function. The objective function used with PPO can be expressed in terms of the probability ratio $p_{k}({\bm \theta})$ given by,
\begin{equation}
\label{eq:clipr} 
p_{k}({\bm \theta})=\frac{\pi_{{\bm \theta}}({\bf u}_{k}|{\bf o}_{k})}{\pi_{{\bm \theta}_\text{old}}({\bf u}_{k}|{\bf o}_{k})}
\end{equation}

The PPO objective function is shown in Equations ~\eqref{eq:ppoloss_a} through ~\eqref{eq:ppoloss_c}.  The general idea is to create two surrogate objectives, the first being the probability ratio $p_{k}({\bm \theta})$ multiplied by the advantages $A^{\pi}_{\bf w}({\bf o}_{k},{\bf u}_{k})$ (see Eq. \eqref{eq:ppo_adv}), and the second a clipped (using clipping parameter $\epsilon$) version of $p_{k}({\bm \theta})$ multiplied by $A^{\pi}_{\bf w}({\bf o}_{k},{\bf u}_{k})$.  The objective to be maximized $J({\bm \theta})$ is then the expectation under the trajectories induced by the policy of the lesser of these two surrogate objectives.

\begin{subequations}
\begin{align}
	\text{obj1} &= p_{k}({\bm \theta})A^{\pi}_{\bf w}({\bf o}_{k},{\bf u}_{k})\label{eq:ppoloss_a}\\
	\text{obj2} &= \mathrm{clip}(p_{k}({\bm \theta})A^{\pi}_{\bf w}({\bf o}_{k},{\bf u}_{k}) , 1-\epsilon, 1+\epsilon)\label{eq:ppoloss_b}\\
	J({\bm \theta})&=\mathbb{E}_{p({\bm \tau})}[\mathrm{min}(\text{obj1},\text{obj2})]\label{eq:ppoloss_c}
\end{align}
\end{subequations}

This clipped objective function has been shown to maintain a bounded Kullback-Leibler (KL) divergence \cite{kullback1951information} with respect to the policy distributions between updates, which aids convergence by ensuring that the policy does not change drastically between updates. Our implementation of PPO uses an approximation to the advantage function that is the difference between the empirical return and a state value function baseline, as shown in Equation \ref{eq:ppo_adv}, where $\gamma$ is a discount rate and $r$ the reward function, described later in Equations \ref{eq:rew1} through \ref{eq:rew5}:
\begin{equation}
\label{eq:ppo_adv}
    A^{\pi}_{\bf w}(\mathbf{x}_{k},\mathbf{u}_{k})=\left[\sum_{\ell=k}^{T}\gamma^{\ell-k}r(\bf o_{\ell},\bf u_{\ell})\right]-V_{\bf w}^{\pi}(\mathbf{x}_{k})
\end{equation}
Here the value function $V_{\bf w}^{\pi}$ is learned using the cost function given by
\begin{equation}
\label{eq:vf_ppo}
L(\mathbf{w})=\sum_{i=1}^{M}\left(V_{\mathbf{w}}^{\pi}({\bf o}_k^i)-\left[\sum_{\ell=k}^{T}\gamma^{\ell-k}r({\bf u}_{\ell}^i,{\bf o}_{\ell}^i)\right]\right)^2
\end{equation}
In practice, policy gradient algorithms update the policy using a batch of trajectories (roll-outs) collected by interaction with the environment. Each trajectory is associated with a single episode, with a sample from a trajectory collected at step $k$ consisting of observation ${\bf o}_{k}$, action ${\bf u}_{k}$, and reward $r_k({\bf o}_k,{\bf u}_k)$. Finally, gradient ascent is performed on ${\bm \theta}$ and gradient descent on ${\bf w}$ and update equations are given by
\begin{align}\label{loss}
{\bf w}^+&={\bf w}^--\beta_{{\bf w}}\nabla_{{\bf w}} \left. L({\bf w})\right|_{{\bf w}={\bf w}^-}\\
{\bm \theta}^+&={\bm \theta}^-+\beta_{{\bm \theta}} \left. \nabla_{\bm \theta}J\left({\bm \theta}\right)\right|_{{\bm \theta}={\bm \theta}^-}
\end{align}
where $\beta_{{\bf w}}$ and $\beta_{{\bm \theta}}$ are the learning rates for the value function, $V_{\bf w}^{\pi}\left({\bf o}_k\right)$, and policy, $\pi_{\bm \theta}\left({\bf u}_k|{\bf o}_k\right)$, respectively.

\subsection{Meta-RL Problem Formulation and Policy Optimization}

In our PPO implementation, we set the clipping parameter $\epsilon$ to 0.1. The policy and value function are learned concurrently, as the estimated value of a state is policy dependent. The policy uses a multi-categorical policy distribution, where a separate observation conditional categorical distribution is maintained for each element of the action vector $\mathbf a \in \mathbb{Z}^{10}$, where in this application, we have 2 possible actions (thruster on / thruster off) for each element of the action vector. Note that because the attitude control thrusters are fired in pairs, the action vector is in $\mathbb{Z}^{10}$ rather than $\mathbb{Z}^{16}$.  Specifically, the action distribution is implemented by applying the softmax function to the two logits (a logit is an output of the policy network) corresponding to each element of the action vector, as shown in Equation \eqref{eq:softmax}, where $p(a_{ij})$ is the probability of taking action $j \in {-1,1}$ for the $i^{th}$ element of the action vector (here $i \in {[0,9]}$) and $z_{ij}$ is the logit corresponding to action $j$. 

\begin{equation}
\label{eq:softmax}
    p(a_{ij} | \mathbf{o}) = \frac{e^{z_{ij} | \mathbf{o}}}{\sum_j e^{z_{ij} | \mathbf{o}} }
\end{equation}

During optimization, the policy samples from this distribution, returning a value in $\mathbb{Z}^{k}$,  Each element of the agent action $\mathbf{u} \in {0,1}$ is used to index Table \ref{tab:thrusters}, where if the  action is 1, it is used to compute the body frame force and torque contributed by that thruster.   For testing and deployment, the sampling is turned off, and the action is just the argmax of the two logits across each element of the action vector. 

Note that exploration is conditioned on the observation, with the two logits associated with each element of the action vector determining how peaked the softmax distribution becomes for each action. Because the probabilities in Equation \ref{eq:clipr} are calculated using the logits, the degree of exploration automatically adapts during learning such that the objective function is maximized. 

We define $\hat{\theta}_{u_o}$ and $\hat{\theta}_{v_o}$ as the stabilized and filtered seeker angles at the start of the homing phase, and the related  signals $\hat{d\theta}_u=\hat{\theta}_u-\hat{\theta}_{u_o}$ and $\hat{d\theta}_v=\hat{\theta}_v-\hat{\theta}_{v_o}$. The observation given to the agent is then as shown in Equation~\eqref{eq:obs}, where $\hat{\mathbf{dq}}$ is defined in Equation \ref{eq:dqi}.

\begin{equation}
    \label{eq:obs}
    \mathrm{obs} = \begin{bmatrix} \hat{d\theta}_u & \hat{d\theta}_u & \hat{\dot{\theta}}_u & \hat{\dot{\theta}}_v & \hat{\mathbf{dq}} & \hat{\boldsymbol{\omega}}  \end{bmatrix} 
\end{equation}

The policy and value functions are implemented using four layer neural networks with tanh activations on each hidden layer. Layer 2 for the policy and value function is a recurrent layer implemented using gated recurrent units \cite{chung2015gated}. The network architectures are as shown in Table \ref{tab:NN}, where $n_{\mathrm{hi}}$ is the number of units in layer $i$, $\mathrm{obs\_dim}$ is the observation dimension, and $\mathrm{act\_dim}$ is the action dimension. The policy and value functions are periodically updated during optimization after accumulating trajectory rollouts of 30 simulated episodes.

\begin{table}[h]
	\fontsize{10}{10}\selectfont
    \caption{Policy and Value Function network architecture}
   \label{tab:NN}
        \centering 
   \newcolumntype{R}{>{\raggedleft\arraybackslash}p{1.8cm}}
   \begin{tabular}{l  R  c  R  c } 
      \hline 
       & \multicolumn{2}{c}{Policy Network} & \multicolumn{2}{c}{Value Network}\\
       \hline
       Layer & \# units & activation & \# units & activation \\
       \hline
      hidden 1      & $10 * \mathrm{obs\_dim}$ & tanh & $10 * \mathrm{obs\_dim}$ & tanh \\
      hidden 2      & $\sqrt{n_{\mathrm{h1}} * n_{\mathrm{h3}}}$ & tanh & $\sqrt{n_{\mathrm{h1}} * n_{\mathrm{h3}}}$ & tanh\\
      hidden 3      & $10 * \mathrm{act\_dim}$ & tanh & 5 & tanh \\
      output        & $\mathrm{act\_dim}$ & linear & 1 & linear \\
      \hline
   \end{tabular}
\end{table}

The agent receives a terminal reward  if the miss distance is less than 50 cm at the end of an episode,  as shown in Equation \ref{eq:rew4}.  Because it is highly unlikely that the agent will experience these rewards through random exploration, we augment the reward function using shaping rewards \cite{ng2003shaping}. These shaping rewards are given to the agent at each timestep, and guide the agent's behavior in such a way that the agent will begin to experience the terminal reward. Specifically, the shaping rewards encourage behavior that  minimizes $\dot{\theta}_u,\dot{\theta}_v$, as shown in Equation~\eqref{eq:rew1}, where $\sigma_{\dot{\theta}}$ is a hyperparameter.  We also encourage minimal use of the attitude control thrusters by penalizing attitude control effort, as shown in Equation \ref{eq:rew2}. We found it was counterproductive to penalize divert thruster firing, as any excessive use of these thrusters was already penalized, as it resulted in larger miss distances. Finally,the agent is rewarded for keeping the missile's attitude $\mathbf{q}$ at the attitude that exists at the start of the homing phase $\mathbf{q}_{\mathrm{init}}$, as shown in Equation \ref{eq:rew3}.  These rewards are then weighted and summed as shown in Equation \ref{eq:rew5}. Reward hyperparameters are summarized in Table \ref{tab:HPS}.

\begin{subequations}
\begin{align}
    r_{\mathrm{shaping}} &= \exp\left(-\frac{\|\begin{bmatrix} \dot{\theta_u} & \dot{\theta_v} \end{bmatrix}\|}{\sigma_{\dot{\theta}}})\right)\label{eq:rew1}\\
    r_{\mathrm{control}} &= \sum_{k=4}^{10}\mathbf{T}_{\mathrm{cmd}}\label{eq:rew2}\\
    r_{\mathrm{attitude}} &= \arccos\left( 2(\mathbf{q}\cdot\mathbf{q}_{\mathrm{init}})^2-1\right)\label{eq:rew3}\\
    r_{\mathrm{terminal}} &=
    \begin{cases}
        1,& \text{if } \mathrm{miss} < 50\mathrm{cm}\\
        0,              & \text{otherwise}
    \end{cases} \label{eq:rew4}\\
    r &= \alpha r_{\mathrm{shaping}} + \beta r_{\mathrm{control}} + \delta r_{\mathrm{attitude}} + \eta r_{\mathrm{terminal}}\label{eq:rew5}
\end{align}
\end{subequations}

\begin{table}[h]
	\fontsize{10}{10}\selectfont
    \caption{Reward Hyperparameter Settings}
   \label{tab:HPS}
        \centering 
   \begin{tabular}{ c  c  c  c  c  } 
      \hline
      $\alpha$  & $\beta$   &  $\delta$ & $\eta$ &  $\sigma_{\dot{\theta}}$\\
      \hline
       1.0 & -0.02 & -0.1 & 10 &  0.04\\
   \end{tabular}
\end{table}

\begin{figure}[h]
\begin{center}
\includegraphics[width=1.0\linewidth]{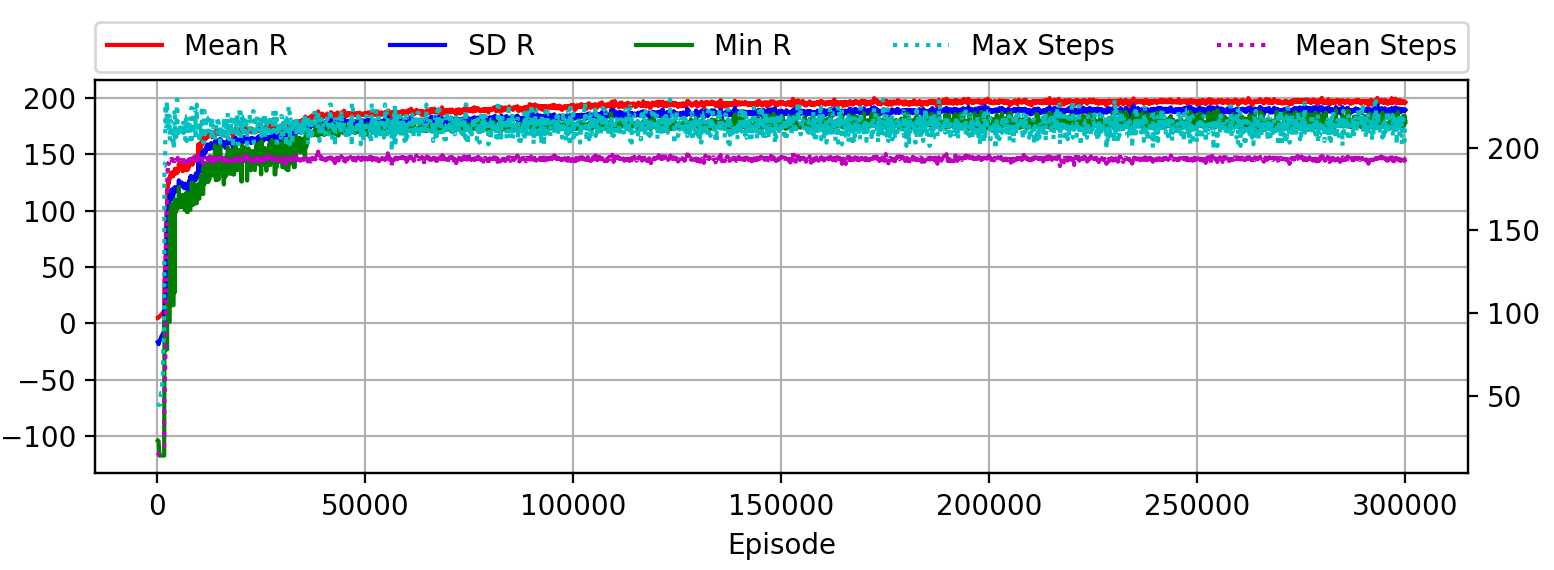}
\caption{Learning Curves: Rewards}
\label{fig:lc_rewards}
\end{center}
\end{figure}

\begin{figure}[h!]
\begin{center}
\includegraphics[width=1.0\linewidth]{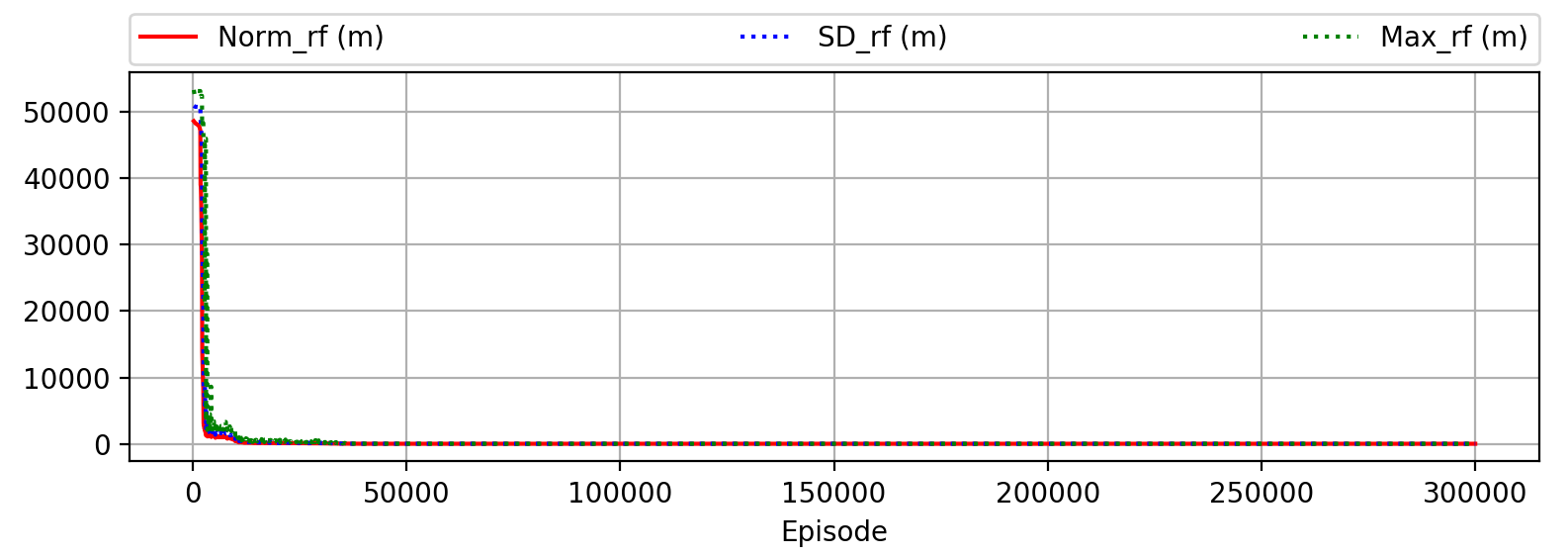}
\caption{Learning Curves: Rewards}
\label{fig:lc_miss}
\end{center}
\end{figure}

During optimization, the policy and value function are updated using rollouts collected over 30 episodes. An episode is terminated if one of the seeker angles $\theta_u, \theta_v$ exceeds half the maximum field of view (90/2=45 degrees). Note that this termination condition covers the cases of successful intercepts and misses, and indirectly implements a field of view constraint. Episode termination also occurs in the case of a rotational velocity constraint violation (any element of the rotational velocity vector exceeds 12 rad/s) or the agent runs out of fuel. We found that optimization was stabilized by setting the missile's dry mass to 10kg for optimization (but we set it back to 25kg for testing).  We use the dual discount rate approach first suggested in \cite{gaudet2018deep}, with shaping rewards discounted by $\gamma_1=0.90$ and terminal rewards discounted by $\gamma_2=0.995$.  Table \ref{tab:Opt_scenario} gives the initial conditions used for optimization where they differ from the values given in \ref{tab:ic}. We did try optimizing with non-zero $e_{\theta}$, $e_{\omega}$, $\sigma_{\theta}$, and $\sigma_{\omega}$, but obtained the best test results when these were set to zero during optimization. Fig. \ref{fig:lc_rewards}  plots the reward statistics for the rewards received by the agent and number of steps per episode over the 30 episodes of rollouts used to update the policy and value function.  Note that a 200 step episode will have a duration of 8s.  Fig.  \ref{fig:lc_miss} gives statistics for miss distance during optimization, again with the statistics calculated over 30 episodes of rollouts. Here the "SD R" curve is the mean reward less one standard deviation. 

\begin{table}[!ht]
    \caption{Optimization Conditions}
   \label{tab:Opt_scenario}
        \centering 
   \begin{tabular}{ l  r  r  r  r  r  r  r } 
      \hline
      Units &  \% & \multicolumn{2}{c}{$\times10^{-3}$} & \multicolumn{2}{c}{$\times10^{-3}$ rad}& \multicolumn{2}{c}{ms}\\
      \hline
      Parameter &  $r_{\mathrm{com}}$ & $e_{\theta}$&  $e_{\omega}$ & $\sigma_{\theta}$ & $\sigma_{\omega}$ & $\tau_{\mathrm{u}}$ & $\tau_{\theta}$ \\
      \hline
      Value & 2.5 & 0.0 & 0.0 & 0.0 & 0.0 & 20 & 20 \\
   \end{tabular}
\end{table}

We did not test different policy and value function architectures or determine the effect of varying the terminal and shaping reward coefficients, and it is possible that performance could be improved with a more extensive hyperparameter search. 

\section{Experiments}\label{Experiments}
Code to reproduce results will be made available at:
\newline
github.com/Aerospace-AI/Aerospace-AI.github.io

\subsection{APN Guidance Law Benchmark}

In the following experiments, we compare the performance of the our integrated GN\&C system to that of the APN guidance law, where we use the zero-effort miss formulation \cite{zarchan2012tactical:6}.  The APN guidance law maps the relative position and velocity between the missile and target and an estimate of the target acceleration to a commanded acceleration in the inertial reference frame, as shown in Equations \eqref{eq:zem1} through \eqref{eq:zem4}, where $N$ is optimally set to 3 for APN. We insert a first order lag to filter the ground truth state and another first order lag to model thruster ignition lag, with the same time constants as used for the Meta-RL policy.  Otherwise, no other parasitic effects are modeled.

\begin{subequations}
\begin{align}
    \mathbf{ZEM} &= \mathbf{r}_\mathrm{TM} + \mathbf{v}_\mathrm{TM}t_{go} + \frac{1}{2}\mathbf{a}_{T}t_{go}^2\label{eq:zem1}\\
    v_c &= -\frac{\mathbf{r}_\mathrm{TM} \cdot \mathbf{v}_\mathrm{TM}}{\|\mathbf{r}_\mathrm{TM}\|}\label{eq:zem2}\\
    t_{go} &= \frac{\|\mathbf{r}_\mathrm{TM}\|}{v_c}\label{eq:zem3}\\
    \mathbf{a}_{\mathrm{com}} &= N \frac{\mathbf{ZEM}}{t_{go}^2}\label{eq:zem4}\\
\end{align}
\end{subequations}

We then convert the commanded acceleration to a body frame acceleration $\mathbf{a}_\mathrm{com}^B$ by projecting it onto the thruster model direction vectors in the first four rows of Table \ref{tab:thrusters}.  Pulsed  thrust is then achieved by turning on an engine when the corresponding element of $\mathbf{a}_\mathrm{com}^B$ exceeds 1/3 of the maximum acceleration (the ratio of maximum thrust to  mass).  This approach allows a closer comparison of the two guidance laws. Moreover, we use the same 6-DOF simulator for testing the APN guidance law and the Meta-RL policy, but with no parasitic effects and no attitude control thrusters, motion is constrained to 3-DOF for the APN case. 

The APN policy was tested against the engagement scenario described in Section \ref{engagement} by running 5000 simulations. It turns out that the APN policy's performance was worse than that of non-augmented PN. This could be due to some combination of the effects of using pulsed thrusters, the sensor and actuator lag, and the types of target maneuvers. Consequently, in the following, we compare the Meta-RL policy to a PN benchmark, with results given in Table~\ref{tab:APN}. 

\begin{table}[!ht]
    \caption{PN performance}
   \label{tab:APN}
        \centering 
   \begin{tabular}{r  r  r  r  r } 
      \hline
      \multicolumn{2}{c}{Miss (cm)} & \multicolumn{3}{c}{Fuel (kg)}\\
      \hline
      $<$ 100 cm (\%) & $<$ 50 cm (\%) & $\mu$ &  $\sigma$ & Max\\
      \hline
      100 & 99 & 8.0 & 1.89 & 18.4 \\
   \end{tabular}
\end{table}

\subsection{Performance of the Integrated GN\&C System}

Table \ref{tab:Scenarios} describes several scenarios used to test our integrated GN\&C system with the optimized Meta-RL policy. Note that since each episode's engagement parameters are randomized at the start of each episode, we are assured that during testing, the agent experiences novel engagement scenarios not seen during optimization. The values in Table \ref{tab:Scenarios} override  the appropriate rows in Table \ref{tab:ic}, otherwise, all other initial conditions are as given in Table \ref{tab:ic}.   For $\sigma_{\theta}$  and $\sigma_{\omega}$ a value of 1 mrad would correspond to a range from -1 mrad to 1 mrad in Table \ref{tab:ic}. Scenario 1 is similar to that of the APN benchmark (Table \ref{tab:APN}), in that the only parasitic effects are $\tau_{\mathrm{u}}$ and $\tau_{\theta}$.

\definecolor{Gray}{gray}{0.9}
\begin{table}[h!]
    \caption{Scenarios}
   \label{tab:Scenarios}
        \centering 
   \begin{tabular}{l  r  r  r  r  r  r  r } 
      \hline
      Units &  \% & \multicolumn{2}{c}{$\times10^{-3}$} & \multicolumn{2}{c}{$\times10^{-3}$ rad} & \multicolumn{2}{c}{ms}\\
      \hline
      Case &  $r_{\mathrm{com}}$ & $e_{\theta}$ & $e_{\omega}$ & $\sigma_{\theta}$ & $\sigma_{\omega}$ & $\tau_{\mathrm{u}}$ & $\tau_{\theta}$ \\
      \hline
      \multicolumn{8}{c}{Similar to PN Test Conditions}\\
      \hline
      1   & 0.0 & 0.0 & 0.0 & 0.0 & 0.0 & 20 & 20 \\
      \hline
      \multicolumn{8}{c}{2.5\% Center of Mass Variation}\\
      \hline
      2  & 2.5 & 0.1 & 0.1 & 0.1 & 0.1 & 20 & 20 \\
      \rowcolor{Gray}
      3  & 2.5 & 1.0 & 1.0 & 1.0 & 1.0 & 20 & 20 \\
      \hline
      \multicolumn{8}{c}{4\% Center of Mass Variation}\\
      \hline
      4  &  4.0 & 0.1 & 0.1 & 0.1 & 0.1 & 20 & 20 \\
      5  &  4.0 & 1.0 & 1.0 & 1.0 & 1.0 & 20 & 20 \\
      \hline
      \multicolumn{8}{c}{First Order Lag Time Constant Variation}\\
      \hline
      6  & 2.5 & 1.0 & 1.0 & 1.0 & 1.0 & 0 & 0 \\
      7  & 2.5 & 1.0 & 1.0 & 1.0 & 1.0 & 10 & 10 \\
      8  & 2.5 & 1.0 & 1.0 & 1.0 & 1.0 & 30 & 30 \\
      \hline
   \end{tabular}
\end{table}

Table \ref{tab:Results} gives the results from running 5000 simulations using randomized initial conditions as shown in Table \ref{tab:ic}, and the parasitic effects tabulated in Table \ref{tab:Scenarios}. In order to focus on performance (as opposed to system component sizing) we set the dry mass to 10 kg for all simulations and report maximum fuel consumption. For scenario 1, where the meta-RL policy has the same parasitic effects as the PN policy (only actuator and sensor lag), we see that performance is almost identical. Scenarios 2 and 3 model the parasitic attitude loop, and we find that performance starts to degrade for scenario 3. When $e_{\theta}, e_{\omega}, \sigma_{\omega}, \sigma_{\omega}$ are further increased above $2\times10^{-3}$, the missile occasionally runs out of fuel and performance falls off steeply. Scenarios 4 and 5 increase the center of mass variation to 4\%. Scenario 5 resulted in rare cases where the missile ran out of fuel, but otherwise increasing the maximum center of mass variation had little impact on performance. 

Scenarios 6-8 investigate the impact of time constants in the guidance system. In endoatmospheric missiles, increasing these time constants attenuates the parasitic attitude loop.  We see a similar effect here when we set $\tau_{\mathrm{u}}$ and $\tau_{\theta}$ to zero (in the simulator, we removed the 1st order lags here to avoid a singularity in the dynamics), accuracy is slightly worse.  But increasing $\tau_{\mathrm{u}}$ and $\tau_{\theta}$ to 30 ms also worsens performance, with the sluggishness impacting accuracy. The impact on fuel efficiency is what we would expect, with higher values of $\tau_{\mathrm{u}}$ and $\tau_{\theta}$ improving fuel efficiency and lower values reducing fuel efficiency. The link between $\tau_{\mathrm{u}}$, $\tau_{\theta}$ and  fuel efficiency comes from the control effort expended trying to track false changes in seeker angles induced by missile rotations in the parasitic attitude loop (see Section \ref{PAL}), with higher time constants damping the missile's response. We highlight row 3 in Tables \ref{tab:Scenarios} and \ref{tab:Results} as we use these conditions in sections \ref{generalization} and \ref{unmodeled}.

\begin{table}[h!]
    \caption{Performance}
   \label{tab:Results}
        \centering 
   \begin{tabular}{l  r  r  r  r  r } 
      \hline
       & \multicolumn{2}{c}{Miss (cm)} & \multicolumn{3}{c}{Fuel (kg)}\\
      \hline
      Scenario & $<$ 100 (\%) & $<$ 50 (\%) & $\mu$ &  $\sigma$ & Max\\
      \hline
      \multicolumn{6}{c}{Similar to PN Test Conditions}\\
      \hline
      1  &  100 & 98 & 9.0 & 3.0 & 19.4 \\
      \hline
      \multicolumn{6}{c}{2.5\% Center of Mass Variation}\\
      \hline
      2 & 100 & 98 & 10.0 & 3.1 & 19.6 \\
      \rowcolor{Gray}
      3  & 99 & 93 & 15.6 & 2.2 & 23.9 \\
      \hline
      \multicolumn{6}{c}{4\% Center of Mass Variation}\\
      \hline
      4  &  100 & 98 & 10.2 & 3.2 & 21.4 \\
      5  &  97 & 89 & 15.9 & 2.5  & 30.1\\
      \hline
      \multicolumn{6}{c}{First Order Lag Time Constant Variation}\\
      \hline
      6 & 99 & 88 & 21.2 & 1.5 & 28.5\\
      7 & 100 & 91 & 16.2 & 2.1 & 23.4\\
      8 & 100 & 92  &  13.8 & 2.6 & 24.0\\
      \hline
   \end{tabular}
\end{table}

\begin{figure}[h!]
\begin{center}
\includegraphics[width=.70\linewidth]{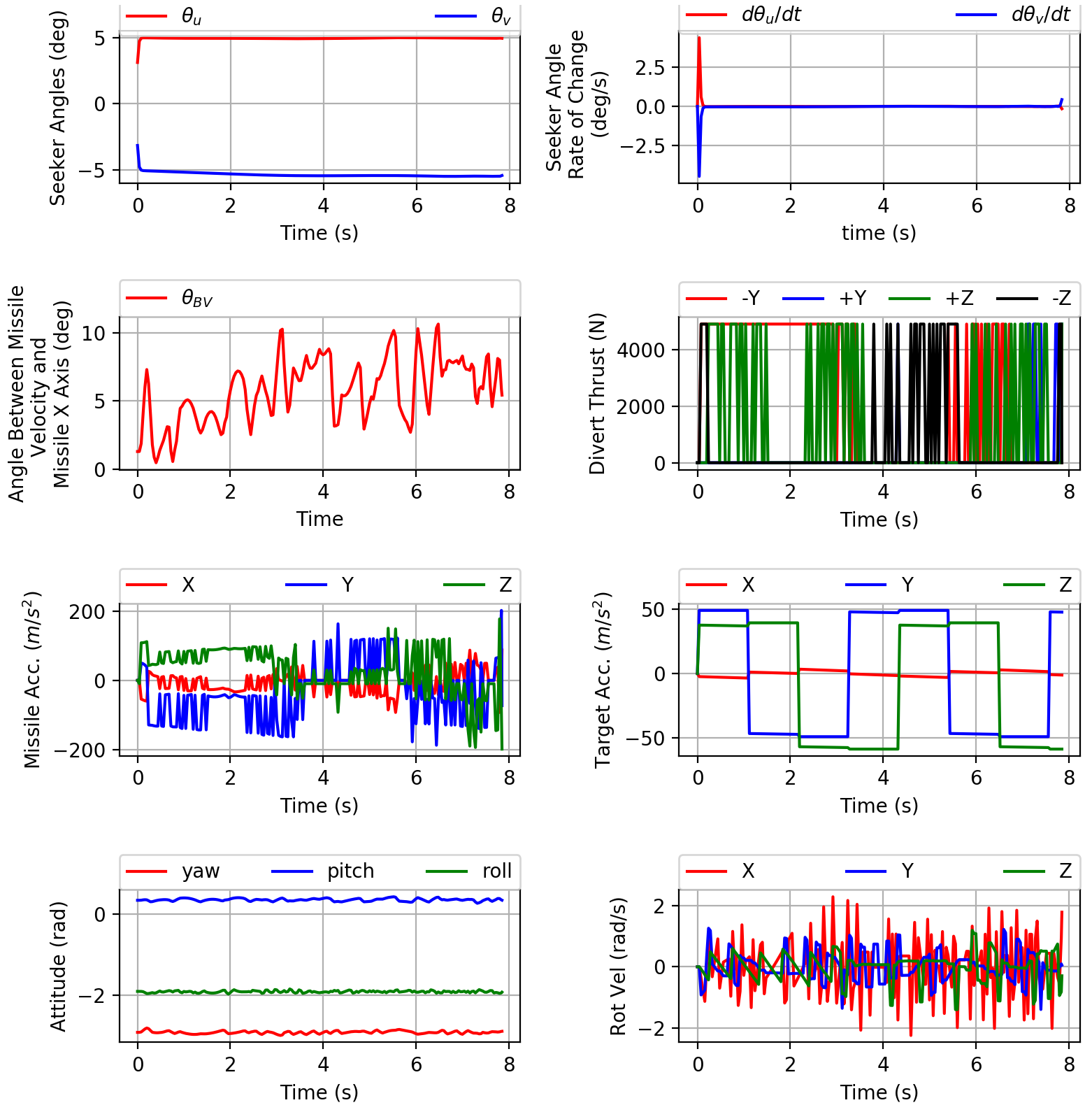}
\caption{Sample Trajectory}
\label{fig:rl_traj}
\end{center}
\end{figure}

\begin{figure}[h!]
\begin{center}
\includegraphics[width=.70\linewidth]{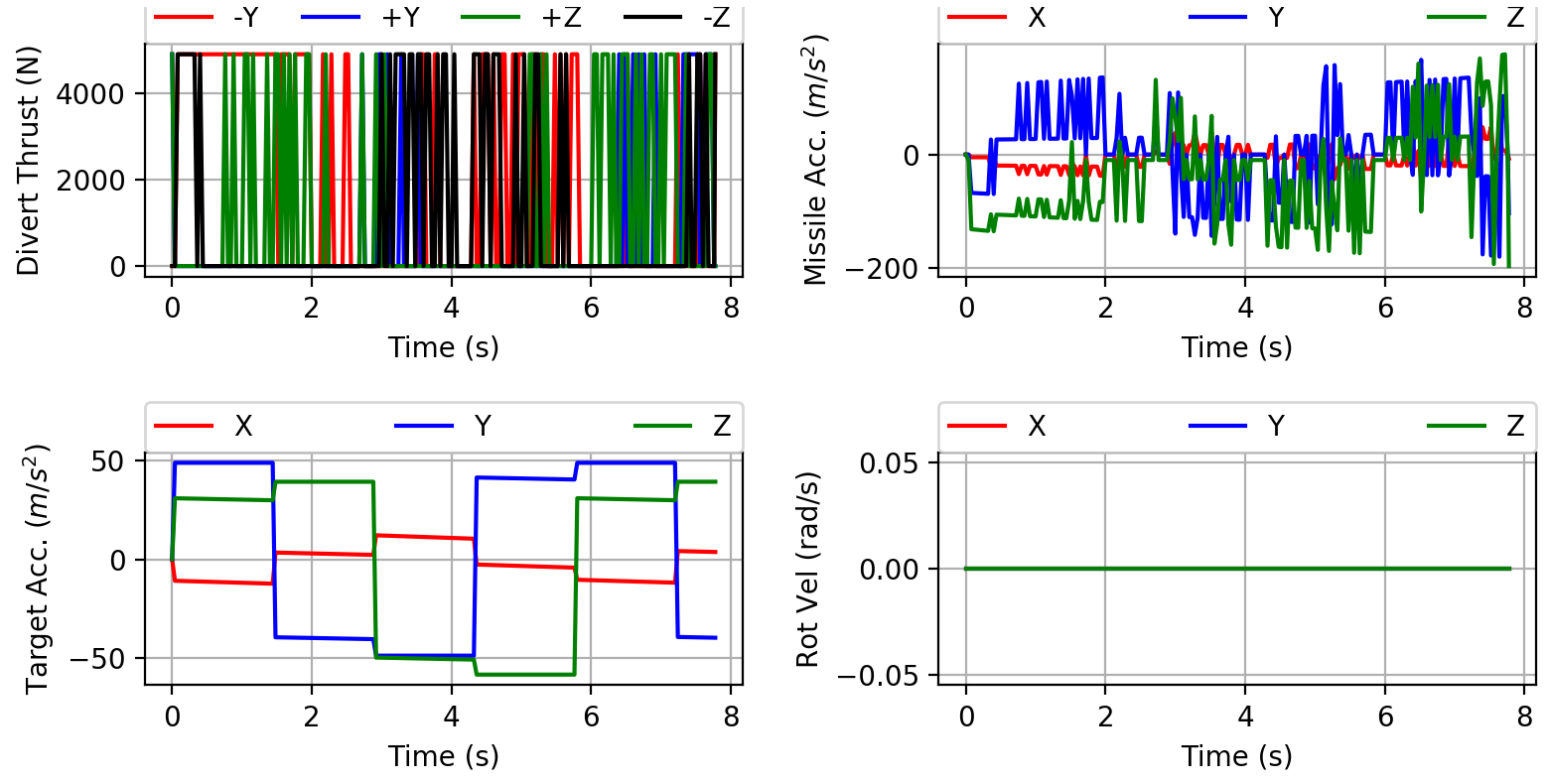}
\caption{Sample Trajectory: No center of mass variation}
\label{fig:rl_traj_nocom}
\end{center}
\end{figure}

\begin{figure}[h!]
\begin{center}
\includegraphics[width=.70\linewidth]{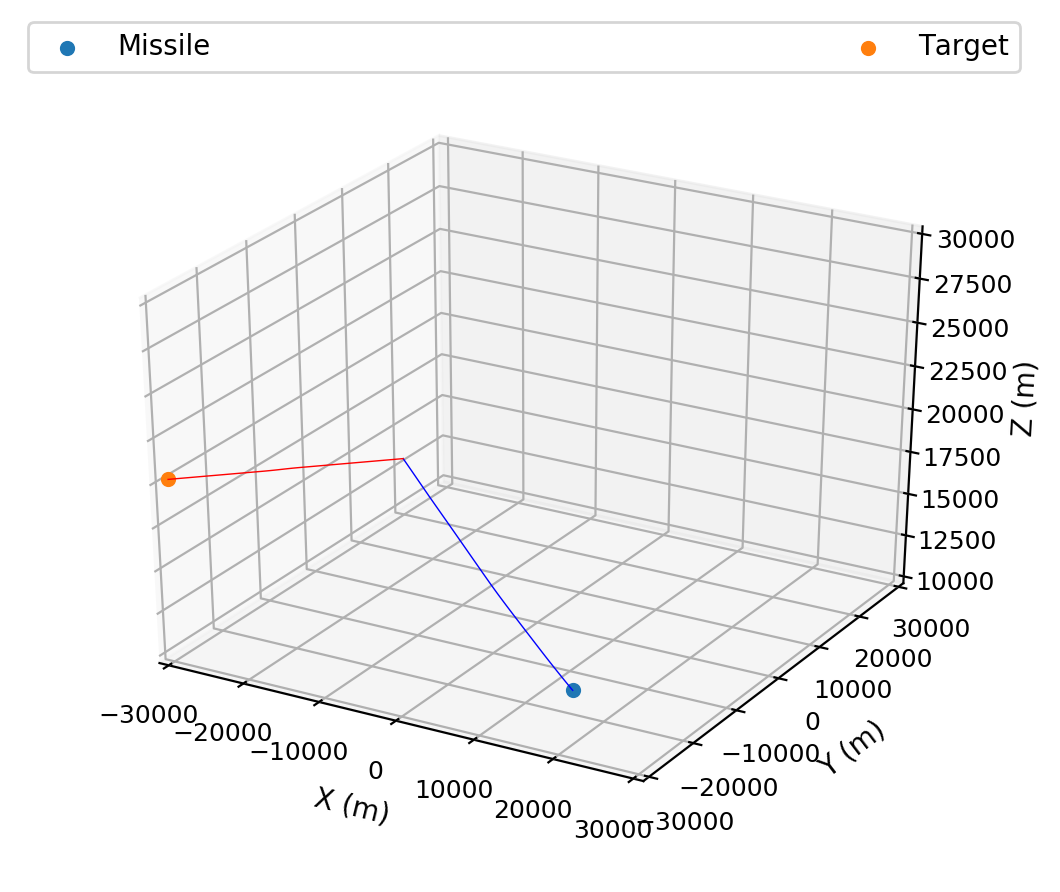}
\caption{Sample 3-D Engagement Plot}
\label{fig:engagement_3d}
\end{center}
\end{figure}

Figure \ref{fig:rl_traj} illustrates a trajectory (scenario 3) with maximum heading error and target acceleration that resulted in a 0.34m miss.  The subplot in the second row, second column, plots $\theta_{BV}$, the angle between the missile's velocity vector and body frame x-axis. Note that even for the case where $\theta_{BV}$ starts at zero, missile divert and attitude control thrusts will in general cause a misalignment between these vectors.   In the top row, we see an abrupt movement in the seeker angles and their rate of change.  This effect disappears when we remove the seeker angle filter and the thruster ignition lag filter, and we speculate that it is due to the initial condition of zero used when integrating the equations of motion simulating the smoothing low pass filter for seeker angles and the first order lag used to model thruster ignition delay. Note that the missile appears to be accelerating in the wrong direction early in the engagement.  We looked at additional plots and found this occurs often, and is likely due to the system attempting to correct for heading error early in the engagement, and only reacting to target maneuvers later. We also found that the system responds less to higher frequency maneuvers, and speculate that this is a learned behavior that improves performance.  We show an additional plot in Figure \ref{fig:rl_traj_nocom} where the center of mass variation is set to zero; this demonstrates that the policy has learned to only make attitude corrections when necessary. Finally, Figure \ref{fig:engagement_3d} illustrates the missile and target positions during a sample engagement. Due to the high closing velocity, the target maneuver and the effects of gravity are not readily apparent in this figure.

\subsection{Generalization to Novel Target Maneuver and Engagement Location}\label{generalization}
 
Here we test the system on a novel target maneuver not seen during optimization, a  barrel roll maneuver \cite{zarchan2012tactical:3} with randomized weave period ranging from 1s to 5s, and the magnitude of target acceleration randomized between zero and the target's maximum acceleration capability. Performance was similar to that of scenario 3 and the results are shown in row 2 of Table \ref{tab:generalization}.  
 
Next, we confirm that the actual location of the engagement in Earth centered coordinates does not impact performance by changing the location to a latitude and longitude of zero, and an elevation of 1000km; results are given in row 3 of Table \ref{tab:generalization}.  To be clear, the engagements corresponding to rows 2 and 3 of Table \ref{tab:generalization} have initial conditions identical to that of scenario 3 except as noted.

\begin{table}[h!]
	\fontsize{10}{10}\selectfont
    \caption{Generalization Results}
   \label{tab:generalization}
        \centering 
   \begin{tabular}{l  r  r  r  r  r } 
      \hline
      Parameter & \multicolumn{2}{c}{Miss (cm)} & \multicolumn{3}{c}{Fuel (kg)}\\
      \hline
      Value & $<$ 100 (\%) & $<$ 50 (\%) & $\mu$ &  $\sigma$ & Max\\
      \hline
      Scenario 3 & 99 & 93 & 15.6 & 2.2 & 23.9 \\
      Weave RL   & 100 & 93 & 15.5 & 2.2 & 23.6 \\
      Equator & 100 & 93 & 15.4 &  2.2 & 23.0 \\
      \hline
   \end{tabular}
\end{table}

\subsection{Extended Engagement Parameters}\label{ext_IC}

To explore the impact of extended engagement parameters on missile performance, we use the initial conditions corresponding to Scenario 2, except we increase the target position angle $\theta$ described in Equations \ref{eq:eng1} through \ref{eq:eng3}. This increases the difference between missile and target flight path angles, which in turn increases the initial angle between the body frame line of sight to target and the seeker boresight axis. Consequently, these engagement parameters will result in the missile's seeker angles taking larger values than experienced in optimization. Intuitively, we expect good generalization if the engagement parameters result in seeker angles and their rates of change remaining within the ranges experienced during optimization. Since this is no longer the case, we expect some deterioration in performance. Indeed, Table \ref{tab:ext_ic} shows that performance slowly falls off as the target position angle is increased, and in some cases the missile uses more than 25kg of fuel. For comparison, the first row of Table \ref{tab:ext_ic} gives the performance for Scenario 2 from Table \ref{tab:Scenarios}. It is worth noting that if we had optimized over a larger range of engagement parameters, performance would have likely been stable over the entire range of scenarios given in Table \ref{tab:ext_ic}. Note that these extended engagement parameters occasionally resulted in infeasible engagements (i.e., a collision triangle did not exist), so we modified our initial condition generator to detect this and keep generating new random engagements until one became feasible.  

\begin{table}[h!]
	\fontsize{10}{10}\selectfont
    \caption{Extended Initial Conditions}
   \label{tab:ext_ic}
        \centering 
   \begin{tabular}{r  r  r  r  r  r  r  r  r } 
      \hline
      \multicolumn{2}{c}{Range (km)} & \multicolumn{2}{c}{Target $\theta$ (deg)} & \multicolumn{2}{c}{Miss (cm)} & \multicolumn{3}{c}{Fuel (kg)}\\
      \hline
      $r_{min}$ & $r_{max}$ & $\theta_{\mathrm{min}}$ & $\theta_{\mathrm{max}}$ & $<$ 100 (\%) & $<$ 50 (\%) & $\mu$ &  $\sigma$ & Max\\
      \hline
      \rowcolor{Gray}
      50 & 55 & 80 & 110 & 100 & 98 & 10.0 & 3.1 & 19.6  \\
      50 & 55 & 100 & 120 & 100 & 97 & 10.9 & 3.4 & 23.7 \\
      50 & 55 & 110 & 130 & 99 & 96 & 11.2 & 3.6 & 27.1 \\
      50 & 55 & 120 & 140 & 98 & 94 & 11.4 &  3.9 & 28.0 \\
      30 & 55 & 80 & 110 & 99 & 97 & 9.6 & 2.0 & 20.1 \\
      30 & 55 & 100 & 120 & 100 & 97 & 10.3 & 2.9  &  22.5\\
      30 & 55 & 110 & 130 & 99 & 95 & 10.8 & 3.6  &  25.4\\
      \hline
   \end{tabular}
\end{table}

\subsection{Policy Adaptation to Simulator Inaccuracy}\label{unmodeled}

A common problem that occurs in engineering is that the system is optimized and tested with the highest fidelity simulator available, but the system still fails during deployment.  These failures can occur due to modeling errors in the missile, environmental dynamics, or both. Here we look at the impact of modeling errors in the missile.  In the following, we simulate these modeling errors using the initial conditions for scenario 3. First, we create a very crude model of fuel slosh, where instead of randomly setting the center of mass variation direction vector once at the start of the episode, we randomly generate it at every simulation time step. The idea is that each divert thrust will result in fuel movement, which will oscillate but decay over time. With many divert thrusts occurring in a typical engagement against a maneuvering target, we just approximate this by randomly shifting the center of mass at each time step.  In row 2 of Table \ref{tab:unmodeled}, we see that accuracy is slightly degraded as compared to scenario 3.

Next, we randomly perturb the missile's nominal inertia tensor (Equation \ref{eq:inertia_tensor}) at the start of each episode. A 20\% perturbation means we perturb the diagonal elements by a uniformly distributed value in the range 20\% of their nominal values and add a uniformly distributed value in the range [-0.2,0.2] for the off-diagonal elements. For the 40\% case the range for the off-diagonal elements is [-0.4,0.4]. Interestingly, we find that this perturbation gives a slight boost to fuel efficiency. We speculate that some of these inertia tensor mutations resulted in a design that was more controllable, and the policy adapted to take advantage of it.

Finally, we look at the case where the thrusters are mismatched in thrust capability. Specifically, at the start of each episode we randomly select a thruster and set its maximum thrust to between 0.80 and 1.00 of nominal thrust capability. Again, the results are very close to that for scenario 3.

 \begin{table}[!ht]
    \caption{Meta-RL Performance with Simulator Inaccuracy}
   \label{tab:unmodeled}
        \centering 
   \begin{tabular}{l  r  r  r  r  r} 
      \hline
      Case & \multicolumn{2}{c}{Miss (cm)} & \multicolumn{3}{c}{Fuel (kg)}\\
      \hline
      Value & $<$ 100 (\%) & $<$ 50 (\%) & $\mu$ &  $\sigma$ & Max\\
      \hline
      Scenario 3 &  99 & 93 & 15.6 & 2.2 & 23.9 \\
      Fuel Slosh  & 100 & 91 & 15.7 & 9.7 & 23.7\\
      20\% ITV & 100 & 93  & 14.3 & 2.4 &  22.4 \\
      40\% ITV & 100 & 93 & 14.5 & 2.5 & 23.9 \\
      Thruster Mismatch & 100 &  92 & 15.6 & 2.2 & 23.6\\
      \hline
   \end{tabular}
\end{table}

\subsection{Discussion}

It appears that the Meta-RL optimized GN\&C system performs reasonably well in the presence of parasitic effects. Performance deteriorates as compared to the 3-DOF APN policy, but still has a high probability of intercept. In the experiments, we find that the parasitic attitude loop has the dominant impact on performance as compared to center of mass variation. At first glance, it might seem that having performance fall off when seeker angle and rotational velocity noise scale factor errors exceed $1\times10^{-3}$ is a problem, as modern endoatmospheric missiles routinely intercept maneuvering targets using radar seekers protected by radomes that introduce scale factor errors as high as $1\times10^{-1}$ \cite{siouris2004missile:2}.  However, in order to reduce the impact of the parasitic attitude loop, relatively high guidance system time constants must be used \cite{zarchan2012tactical:5}, which would result in large misses for the more demanding exoatmospheric interception scenario. To illustrate, Zarachan simulates an endoatmospheric engagement with a radome slope of $-1\times10^{-2}$, and uses a 0.5 s guidance system time constant in order to stabilize the parasitic attitude loop. Similarly, Willman \cite{willman1988effects} uses a 2s guidance system time constant with a 0.04 scale factor error. But a 0.5 s guidance system time constant would result in large miss distances in the exoatmospheric applications, as we confirmed by running the PN policy with $\tau_{\theta}=0.1\text{ s}$ and $\tau_{\theta}=0.5\text{ s}$ in the scenario from row 1 of Table \ref{tab:APN}, with the probability of a successful intercept falling to 70\% and  to 0\% respectively. That said, tolerance to Gaussian noise could likely be improved by using a Kalman filter to provide better estimates of the rotational velocity and seeker angle measurements, although the goal of this work was an integrated GN\&C system, with the policy acting on minimally  processed raw sensor inputs. Moreover, it appears that modern rate gyros have scale factor accuracy of around 200ppm \cite{sun2015line}, in which case our results appear satisfactory.

One question that remains is how much performance is attributable to the policy adapting to environmental variation as compared to simple robustness.  In previous work \cite{gaudet2019adaptive} we were able to address this by using two policies, one with a recurrent network layer than adapts to the environment and another  using only fully connected network layers. In this case the improved performance of the recurrent policy was attributed to adaptation. However, in this work the policy observations do not include range and closing velocity, and consequently does not satisfy the Markov property, i.e., we have a POMPD. It follows that a recurrent policy that can capture the history of an engagement in a hidden state is required to solve the problem, and it is therefore not possible to compare a non-recurrent and recurrent policy.

\section{Conclusion}

We optimized an adaptive integrated guidance, navigation, and control system using Reinforcement Meta-Learning in a six degrees-of-freedom environment. The system implements minimal processing of sensor outputs: a seeker angle stabilization scheme that works with coupled yaw and pitch channels, a simple lag filter to smooth the seeker angles, and an integrator to estimate change in attitude. A Meta-RL optimized policy then maps the stabilized and filtered seeker angles, their rates of change, the estimated change in attitude since the start of the homing phase, and the estimated rotational velocity directly to thruster on/off commands. The system is optimized and tested in an environment that models parasitic effects including seeker angle lag and thruster ignition delay, center of mass variation from consuming fuel, and scale factor errors and Gaussian noise on sensor outputs, the latter two creating a parasitic attitude loop. This appears to be the first published work modeling the parasitic attitude loop in a 6-DOF exoatmospheric intercept that takes into account the interplay between center of mass variation induced by divert thrusts and scale factor errors on the seeker angles and rotational velocity.  We also believe that we have demonstrated a higher level of integration (guidance, navigation, and control) in the system being optimized than previous work. The optimized system was found to be reasonably robust to parasitic effects, with performance close to that of a three degrees-of-freedom augmented proportional navigation system with perfect knowledge of the engagement state and no parasitic attitude loop. The system generalizes well to novel engagement scenarios not experienced during optimization, including an extended initial condition range and novel target maneuvers. The system also performed well when confronted with simulation model inaccuracy, including unmodeled fuel slosh, inertia tensor variation, and thruster mismatch. Future work will integrate target discrimination, and explore integrated guidance, navigation, and control for the endoatmospheric intercept problem.

\bibliography{references}

\end{document}